\documentclass[prl,showpacs,preprintnumbers,twocolumn,amsmath,amssymb,superscriptaddress]{revtex4}
\usepackage{graphicx}
\usepackage{color}
\usepackage{dcolumn}
\usepackage{bm}
\usepackage{psfrag}
\def\bq{{\bf q}}

\newcommand{\be}{\begin{equation}}
\newcommand{\ee}{\end{equation}}
\newcommand{\bea}{\begin{eqnarray}}
\newcommand{\eea}{\end{eqnarray}}

\def\a{\alpha}

\def\e{\varepsilon}
\def\d{\delta}

\def\l{\lambda}

\renewcommand{\o}{\omega}

\def\s{\sigma}

\def\D{\Delta}

\def\ra{\rightarrow}

\def\pd{\partial}
\def\nb{\nabla}
\def\bk{{\bf k}}

\def\bq{{\bf q}}

\def\bQ{{\bf Q}}
\def\bA{{\bf A}}

\def\bJ{{\bf J}}

\def\bx{{\bf x}}

\def\nn{\nonumber}
\def\lb{\label}

\def\pref#1{(\ref{#1})}

\newcount\bozza \bozza=0
\ifnum\bozza=1
\newdimen\shift \shift=-2truecm
\def\lb#1{%
{\label{#1}\rlap{\kern\shift{$\scriptstyle#1$}}}}
\else\def\lb#1{\label{#1}} \fi

\begin{document}
\title{Non-relativistic dynamics of the amplitude (Higgs) mode in superconductors}

\author{T. Cea}
\affiliation{ISC-CNR and Dep. of Physics, ``Sapienza'' University of
  Rome, P.le A. Moro 5, 00185, Rome, Italy}
\email{lara.benfatto@roma1.infn.it}
\author{C. Castellani}
\affiliation{ISC-CNR and Dep. of Physics, ``Sapienza'' University of
  Rome, P.le A. Moro 5, 00185, Rome, Italy}
\author{G. Seibold}
\affiliation{Institut F\"ur Physik, BTU Cottbus-Senftenberg, PBox 101344, 03013 Cottbus, Germany}
\author{L. Benfatto}
\affiliation{ISC-CNR and Dep. of Physics, ``Sapienza'' University of
  Rome, P.le A. Moro 5, 00185, Rome, Italy}

\date{\today}

\begin{abstract}

Despite the formal analogy with the Higgs particle, the amplitude fluctuations of the order parameter in weakly-coupled superconductors do not identify a real mode with a Lorentz-invariant dynamics. Indeed, its resonance occurs at $2\D_0$, which coincides with the threshold $2 E_{gap}$ for quasiparticle excitations, that spoil any relativistic dynamics. Here we investigate the fate of the Higgs mode in the unconventional case where $2 E_{gap}$ becomes larger than $2\D_0$, as due to strong coupling or strong disorder.  We show that also in this situation the amplitude fluctuations never identify a real mode at $2\D_0$, since such  "bosonic" limit is always reached via a strong mixing with the phase fluctuations, which dominate the low-energy part of the spectrum. Our results have direct implications for the interpretation of the sub-gap optical absorption in disordered superconductors. 


\end{abstract}


\maketitle

In the BCS theory of superconductivity the formation of Cooper pairs and their condensation into the superfluid state are both consequences of the 
spontaneous breaking of the $U(1)$ gauge symmetry at $T_c$\cite{nagaosa}. While the equilibrium value of the  (complex) order parameter $\D_0$ is responsible for a gap in the single-particle excitation spectrum,  the fluctuations of its phase represent the massless Goldstone mode that couples to the electromagnetic field leading to the Meissner effect. In addition, energetically-costly collective amplitude fluctuations are also possible \cite{schmid68,volkov73,kulik81,varma_prb82,sharapov_mgb2,marciani_prb13,nikuni_prb13,shimano14,cea_cdw_prb14,randeria14,varma_review,anderson_comm15, frydman_natphys15,deveraux_cm14}. The resulting collective mode is sometimes named in the literature "Schmid mode"\cite{schmid68,kulik81,sharapov_mgb2,marciani_prb13}, or alternatively "Higgs mode"\cite{
nikuni_prb13,shimano14,cea_cdw_prb14,randeria14,varma_review,anderson_comm15, frydman_natphys15,deveraux_cm14}  to emphasise its analogy with the massive particle of the Standard Model\cite{weinberg}.

However, such analogy holds only in part, since the dynamics of the superconducting (SC) Higgs mode is not described in general by a Lorentz-invariant (LI) relativistic theory. Indeed, in conventional  weakly-coupled superconductors the Higgs resonance occurs exactly at the threshold $2\Delta_0$ for the proliferation of quasiparticle pairs that fully control its dynamics leading to a non-relativistic strongly overdamped mode\cite{kulik81,varma_prb82,cea_cdw_prb14},  whose experimental signature emerges usually only in out-of-equilbrium spectroscopy\cite{shimano,carbone,shimano14,deveraux_cm14}. 
The situation can be different e.g. for lattice bosons at commensurate fillings, where fermionic quasiparticle excitations are absent. In this case the system has been described\cite{fisher_prb89,sachdev_prb97,altman_prl08,auerbach_prb10,podolsky_prb11,prokofev_prl12,podolsky_prl13}  by a SC-like LI $O(2)$ model:
\be
\lb{so2}
S=\int dt d\bx \frac{1}{2}\left[-|\pd_t\psi|^2+c^2|\nb \psi|^2+\frac{m^2}{4\D_0^2}\left(
|\psi|^2-\D_0^2\right)^2\right]
\ee
where $\psi=(\D_0+\Delta) e^{i\theta}$ includes  the amplitude ($\D$) and phase ($\theta$) fluctuations of the SC order parameter.
By retaining in Eq.\ \pref{so2} only Gaussian terms one has:
\be
\lb{paradigm}
\langle |\Delta(q)|^2\rangle=\frac{1}{-\o^2+m^2+c^2\bq^2}, \, 
\langle |\theta(q)|^2\rangle=\frac{\D_0^{-2}}{-\o^2+c^2\bq^2},
\ee
i.e. both the phase (Bogoliubov) and amplitude (Higgs) modes appear as elementary excitations with a LI dispersion. The Higgs resonance, that is not overdamped\cite{podolsky_prb11,prokofev_prl12,podolsky_prl13} by the decay processes into phase modes at higher orders,  has been probed in cold atoms by properly shaking the optical lattice\cite{coldatoms1,coldatoms2}. Motivated by this result, it has been recently suggested\cite{auerbach_prb10,podolsky_prb11,frydman_natphys15} 
that the $O(2)$ model \pref{so2} is also the correct paradigm for "bosonic" superconductors, i.e. fermionic systems where the energetic cost to create single-particle excitations $E_{gap}$  is {\em larger} than the SC order parameter $\Delta_0$. 
The underlying idea is that the if Higgs resonance  remains at $2\Delta_0$ it is untouched by the quasiparticle proliferation at $2E_{gap}$, and the Higgs mode recovers its LI dynamics, as recently suggested for 
charge-density-wave (CDW) superconductors\cite{cea_cdw_prb14}. 
Apart from this peculiar case, in general the separation between $E_{gap}$ and $\Delta_0$ occurs in homogeneous superconductors in the strong-SC coupling limit\cite{review_micnas,bec_randeria}. In addition, it can also be induced  at weak coupling by strong disorder, as it has been observed experimentally\cite{sacepe09,mondal11,sacepe11,chand12,noat13,pratap13} in materials like e.g.  InO$_x$ and NbN, and understood theoretically as the effect of the localization of bosonic pairs with large $E_{gap}$\cite{feigelman10,randeria01,trivedi_natphys11}. 
However, in both cases it has not been proven yet the existence or not of a sharp Higgs mode at $2\D_0$, which is also the prerequisite used recently\cite{frydman_natphys15} to interpret the observed sub-gap optical absorption in films of disordered superconductors\cite{armitage_prb07,frydman_natphys15,bachar_jltp14,practh_cm15}. 

In this Letter we demonstrate the lack of a LI Higgs resonance at $2\D_0$ for fermionic 
superconductors. At weak coupling, where
the amplitude and phase sectors are decoupled because of the particle-hole symmetry of
the BCS solution, this is a well-known\cite{kulik81,varma_prb82,cea_cdw_prb14}
effect of the quasiparticles proliferation at $2\D_0$, which overdamp
the Higgs mode and spoil its
relativistic dynamics. In the bosonic limit, where the
single-particle excitation energy $E_{gap}$ overcomes $\D_0$, 
the quasiparticles are ineffective but amplitude and phase sectors are
strongly coupled, and any signature below $E_{gap}$ in the Higgs mode originates by its mixing to the phase mode. In the homogeneous case, where the regime  $E_{gap}>\D_0$  is achieved  at strong coupling  by unavoidably breaking the particle-hole symmetry, only the (Bogoliubov) sound-like phase mode survives at low energy and long wavelength\cite{sofo_prb92,randeria_prb94,alm_prb96,randeria_prb97,depalo_prb99,ganesh_prb09}. In the disordered case the mixing of the two sectors occurs already for weak coupling. However, the spectral weight found below $2E_{gap}$ in the Higgs spectrum does not identify a LI  sharp resonance at $2\D_0$, questioning the proposed\cite{frydman_natphys15}  relevance of the Higgs mode for the sub-gap optical absorption in disordered SC films.

We start our analysis with the homogeneous case, by using as prototype lattice model for a superconductor the attractive Hubbard model on the square lattice,  
\begin{equation}\lb{H_MODEL}
H=-t\sum_{\langle i,j\rangle\sigma}\left(c^\dagger_{i\sigma}c_{j\sigma}+h.c.\right)-U\sum_ic^\dagger_{i\uparrow}c^\dagger_{i\downarrow}c_{i\downarrow}c_{i\uparrow}
\end{equation}
where $U>0$ is the SC coupling strength and $t$ the nearest-neighbor hopping parameter. By treating the interaction term in mean-field approximation the Green's function in the usual basis of Nambu operators reads $G_0^{-1}=i\o_n\hat\s_0-\xi_\bk\hat\s_3+\D_0\hat\s_1$, where $\hat\s_i$ are Pauli matrices,  $\D_0$ is the SC gap and $\xi_\bk=-2t(\cos k_x+\cos k_y)-\mu$
and $E_\bk=\sqrt{\xi_\bk^2+\D_0^2}$  the quasiparticle excitation in the normal and SC state, respectively ($\mu$ being the chemical potential). 
The spectrum of the collective modes can be studied in the effective-action formalism\cite{nagaosa,benfatto_prb04,suppl}, by decoupling 
the interaction term of the model \pref{H_MODEL} both in the pairing and density channels by means of the Hubbard-Stratonovich transformation.
After integrating out the fermions one is left with $S=S_{MF}+S_{FL}$, where $S_{MF}$ is the mean-field saddle point action and $S_{FL}$ the effective action for the collective degrees of freedom $\rho,\Delta,\theta$, connected to the electron density, and the SC amplitude and phase, respectively\cite{suppl}.  
By retaining only the Gaussian terms in the expansion of $S_{FL}$, and integrating out the density field $\rho$,
the effective action for the SC degrees of freedom\cite{suppl} reads $S_{\D\theta}=\sum_q\Psi^+(q)\hat M \Psi(q)$, where  $\Psi(q)\equiv (\D(q),\theta(q))$ and  $\hat M$ at $T=0$ and $\bq\ra0$ is 
\begin{equation}
\lb{gaussian}
\hat M=		\begin{pmatrix}
	(4\D_0^2-\o^2) F(\o) -\frac{U\chi^2_{\rho\D}\tilde\chi_{\rho\rho}}{2\chi_{\rho\rho}}&&
	\frac{i\o}{2} \frac{\chi_{\rho\D}\tilde\chi_{\rho\rho}}{\chi_{\rho\rho}}\\
	-\frac{i\o}{2} \frac{\chi_{\rho\D}\tilde\chi_{\rho\rho}}{\chi_{\rho\rho}} &&
	\frac{\o^2}{4}\tilde\chi_{\rho\rho}+\frac{D_s}{4}\mathbf{q}^2
	\end{pmatrix}.
	\end{equation}
Here $D_s=\frac{1}{2N}\sum_{\mathbf{k}\nu}\frac{\partial^2\xi_\mathbf{k}}{\partial k_\nu ^2}(1-\xi_\mathbf{k}/E_\mathbf{k})$ is the phase stiffness, 
$\chi_{ij}(q)=(T/N)\sum_{\bk,i\o_n}\mathrm{Tr} \left[ G_0(k)\hat\s_iG_0(k-q)\hat\s_j\right]$ is the BCS response function with the identification  $\hat\s_1 \ra \D$ and $\hat\s_3\ra \rho$, while $\tilde\chi_{\rho\rho}\equiv \chi_{\rho\rho}/(1+(U/2)\chi_{\rho\rho})$ is the one dressed at RPA level by density fluctuations\cite{suppl}. In particular at $\bq=0$ one has $\chi_{\rho\rho}\equiv -4\D_0^2 F(\o)$ and 
\bea
\lb{chi31}
\chi_{\rho\D}(\o)&=&-\frac{1}{N}\sum_\bk \frac{4\D_0\xi_\bk}{{E_\bk(4E_\bk^2-(\o+i\d)^2)}},\\
\lb{chif}
F(\o)&=&\frac{1}{N}\sum_\bk \frac{1}{{E_\bk(4E_\bk^2-(\o+i\d)^2)}}.
\eea
The spectrum of the collective modes is obtained as solution of the equation $|\hat M(\o,\bq)|=0$. This simplifies considerably in the 
weak-coupling limit, where $\chi_{\rho\D}\simeq 0$ because of the intrinsic particle-hole symmetry of the BCS solution\cite{cea_cdw_prb14,varma_prb82,randeria_prb94,randeria_prb97,depalo_prb99}, that enforces the integration over a symmetric energy interval around  $\xi=0$   in Eq.\ \pref{chi31}. 
As a consequence the amplitude and phase sectors are decoupled and the matrix \pref{gaussian} reduces to:
\begin{equation}
\lb{BCS}
\hat M^{BCS}\simeq 		\begin{pmatrix}
	(4\D_0^2-\o^2) F(\o) && 0\\
	0&& 
	-\frac{\o^2}{4}\tilde \kappa+\frac{D_s}{4}\bq^2
	\end{pmatrix}
	\end{equation}
where we also used the result, valid in the BCS limit,  $F(\o\ra 0)= N_F/(4\D_0^2)$, so that $-\tilde\chi_{\rho\rho}\simeq
\tilde \kappa=N_F/(1-UN_F/2)$, $N_F$ being the DOS at the Fermi level. From the $M^{BCS}_{22}\equiv\langle |\theta|^2\rangle^{-1}$ element of the matrix \pref{gaussian} one recovers the Anderson-Bogoliubov sound-like mode, which is the only expected phase-density mode at $T=0$\cite{cg_note}. The vanishing of the $M^{BCS}_{11}\equiv \langle |\D|^2\rangle^{-1}$ element identifies instead the pole of the amplitude fluctuations. As one can easily see from Eq.\ \pref{chif}, as $\o\ra 2\D_0$ is 
\be
F(\o)\simeq \frac{N_F \pi}{2\D_0\sqrt{4\D_0^2-\o^2}},\quad \o<2\D_0
\ee
so that the real part of the inverse Higgs propagator vanishes as $\sqrt{4\D_0^2-\o^2}$ at the Higgs mass $m=2\D_0$, which coincides here with the threshold for two-particle excitations. In addition, $\mathrm {Im}F(\o)\sim ({\o^2-4\D_0^2})^{-1/2}$ at $\o>2\D_0$, leading to the usual overdamping of the Higgs spectral function in the  BCS limit\cite{kulik81,varma_prb82,cea_cdw_prb14}. Unless one considers specific band structures\cite{nikuni_prb13}, the same holds in general also at finite momentum, since the Higgs pole remains inside the quasiparticle continuum.


%
\begin{figure}[htb]
\includegraphics[width=8cm,height=5cm,clip=true]{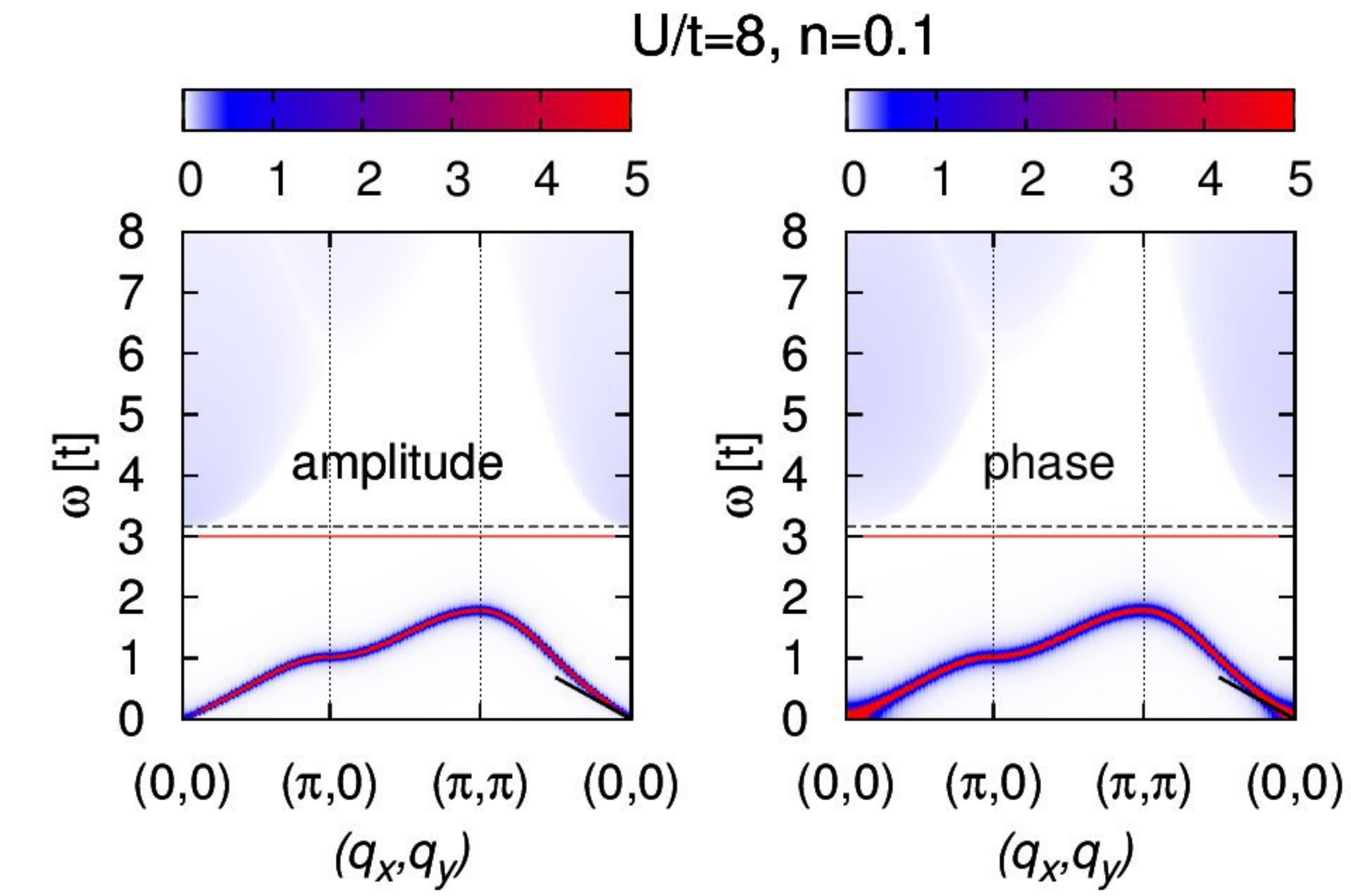}\\
\includegraphics[width=8cm,height=5cm,clip=true]{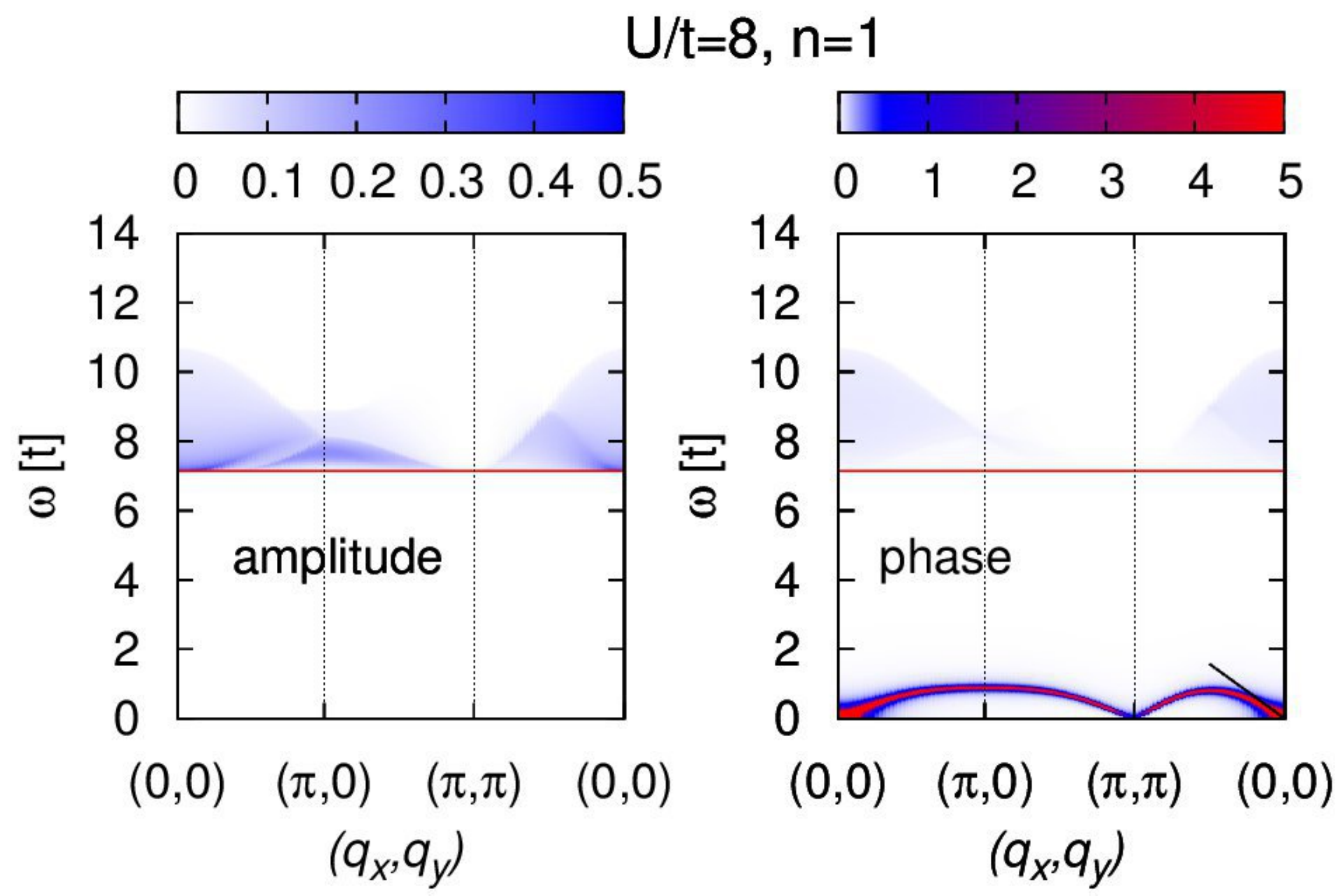}
\caption{Intensity map of the spectral function of the amplitude and phase modes in the homogeneous case at  $U=8t$ for $n=0.1$ (upper panels) and $n=1$ (lower panels). The solid red lines mark the value of $2\Delta_0$ and the dashed black lines the value of $2E_{gap}$. At $n=0.1$ the bosonic limit $E_{gap}>\D_0$ is reached, but the amplitude mode is strongly mixed with the phase, which dominates any sub-gap structure of the Higgs (with zero spectral weight at $\o=0$). At $n=1$ the system is always particle-hole symmetric so the amplitude and phase sectors remain decoupled. However, here $E_{gap}=\D_0$ and the Higgs mode has a broad spectral weight only above $2E_{gap}$. The bending back of the phase mode at $\bQ=(\pi,\pi)$  (where its weight is zero) is a signature of the degenerate CDW instability present in the model \pref{H_MODEL} at $n=1$\cite{benfatto_prb02}.}
\label{fig-u8}
\end{figure}

The BCS result \pref{BCS} shows that even when the particle-hole symmetry leads to a vanishing of the first-order time derivatives\cite{varma_review}, i.e. the off-diagonal terms of Eq.\ \pref{gaussian} and \pref{BCS}, the LI dynamics of the Higgs mode is prevented by the dynamical contribution of quasiparticles, encoded in the singular function $F(\o)$. On the other hand, in the unconventional situation where the quasiparticle continuum $2E_{gap}$ moves away from $2\Delta_0$ one expects that $F(2\D_0)\simeq \mathrm{const}$. Thus, within the diagonal structure  \pref{BCS} for the collective modes one would conclude that the Higgs mode should recover a LI dynamics, as recently 
suggested\cite{auerbach_prb10,podolsky_prb11, frydman_natphys15}. However, in the limit $2E_{gap}\gg2\D_0$ the dynamics of collective modes is no more described by Eq.\ \pref{BCS}. To show this let us first consider the homogeneous strong-coupling limit  $U\gg t$, where  the SC order parameter  and chemical potential for  the model 
\pref{H_MODEL} read\cite{randeria_prb94,benfatto_prb02, suppl}:
\begin{equation}
\lb{sc}
\Delta_0\simeq \frac{U}{2}\sqrt{n(2-n)}, \quad \mu\simeq -\frac{U}{2}(1-n).
\end{equation}
At $n\neq 1$, as soon as the chemical potential goes below the band edge,  $E_{gap}=\sqrt{\D_0^2+(4t-|\mu|)^2}$ becomes larger than $\D_0$. However, 
in this limit the particle-hole symmetry is strongly violated, leading to a large $\chi_{\rho\D}$ mixing between the amplitude and phase sectors in Eq.\ \pref{gaussian}. This can be easily understood from Eq.\ \pref{chi31} in the limit $t/U\simeq 0$, where $|\mu|\gg \D_0$ and $\chi_{\rho\D}\simeq 4\D_0\mu F(\o)$, with $F(\o)\simeq 1/(E(4E^2-\o^2))$ and $E=\sqrt{\D_0^2+\mu^2}\simeq U/2$, see Eq.\ \pref{sc}. In this situation the determinant of the matrix  \pref{gaussian} at $\bq=0$ is given by
\bea
|\hat M|&=&\frac{\o^2}{4}\tilde\chi_{\rho\rho} \left[ (4\D_0^2-\o^2)F(\o)-\frac{\chi_{\rho\D}^2}{\chi_{\rho\rho}}\right]\nn\simeq \\
\lb{strong}
&=&\frac{\o^2}{4}\tilde\chi_{\rho\rho} F(\o)[4(\D_0^2+\mu^2)-\o^2]\simeq \frac{\o^2}{U^2} 
\eea
As one can see, the coupling $\chi_{\rho\D}$ between the amplitude and phase sectors removes completely any signature at $2\D_0$ and the only solution of $|\hat M|=0$ is the (Bogoliubov-like) phase mode\cite{sofo_prb92,randeria_prb94,alm_prb96,randeria_prb97,depalo_prb99,ganesh_prb09}. One can easily see\cite{suppl} that this result holds also in the presence of long-range interactions, that modify the prefactor in the first line of Eq.\ \pref{strong} lifting the phase mode to the plasmon, but do not alter the combination in square brackets, responsible for the disappearance of the pole at $2\Delta_0$. By retaining a finite $t$ value  in the evaluation of all the response functions \pref{chi31}-\pref{chif} one can show\cite{suppl} that the spectral function of the Higgs mode still preserves some small spectral weight above $2E_{gap}$, that is however strongly suppressed with respect to the weak-coupling case. 
These analytical estimates are confirmed by the numerical computation of the amplitude and phase spectral functions\cite{suppl} at finite $\o$ and $\bq$, shown in the upper panels of Fig.\ \ref{fig-u8}. As one can see, any sub-gap feature in the Higgs spectral function is present only at finite $\bq$ and it comes from the mixing to the phase, with no signature at the energy $2\D_0$, marked by the red line. On the other hand at half filling ($n=1$) $\mu=0$ so that 
 it is always $E_{gap}=\D_0$. Moreover, since the  particle-hole symmetry is preserved 
at all orders in $U$, it is  $\chi_{\rho\D}=0$,  the amplitude and phase sectors remain decoupled and the Higgs mode shows only a weak spectral weight  above $2E_{gap}$,  see lower panels of Fig.\ \ref{fig-u8}. Notice that  
results from previous work\cite{seibold_prb04} suggest that RPA describes reasonably well the elementary (Gaussian) collective excitations in the symmetry-broken state at strong coupling. Thus, even though one cannot exclude the presence of additional resonances at higher order, analogous e.g. to that found in the half-filled Bose-Hubbard model very near the superfluid-insulator transition\cite{altman_prl08,auerbach_prb10,prokofev_prl12}, they should not be interpreted as signatures of an elementary amplitude mode.
 

%
\begin{figure}[htb]
\includegraphics[width=9cm,clip=true]{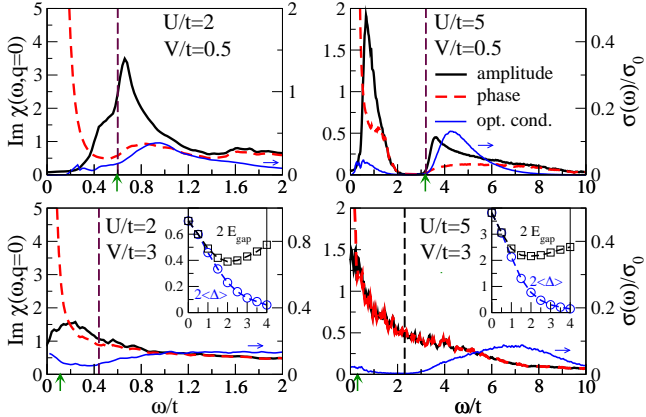}
\caption{Spectral functions of the amplitude and phase modes at $\bq=0$ for $n=0.875$ obtained on a 20$\times$20 lattice after average over 50
disorder configurations for two values of SC coupling ($U/t$) and disorder ($V_0/t$).  The vertical dashed lines and green arrows mark the spectral gap $2E_{gap}$ and $2\langle \D_0\rangle$, respectively, whose disorder dependence is shown in the insets. For comparison we also show $\s(\o)$ (blue line) from Ref.\ \cite{cea_prb14}.}
\label{fig-dis}
\end{figure}
%

A second possible route\cite{randeria01,trivedi_natphys11} to achieve the separation between $E_{gap}$ and $\D_0$ in the model \pref{H_MODEL} is by introducing  disorder as a random on-site energy $V_i$ uniformly distributed in the interval $[-V_0,V_0]$. Indeed, already at the level of the mean-field inhomogeneous Bogoliubov-de-Gennes equations\cite{randeria01} one sees that while the average order parameter $\langle \D_0 \rangle$ decreases as $V_0$ increases, the spectral gap $2E_{gap}$ in the average density of states saturates to a finite value, see insets of Fig.\ \ref{fig-dis}, signaling the formation of local boson pairs\cite{feigelman10,randeria01,trivedi_natphys11}. 
However, once more this does {\em not} imply that the Higgs mode emerges as a sharp resonance at 2$\langle \D_0\rangle$, but instead it acquires sub-gap spectral weight due to the mixing to the phase mode, that is induced in the disordered case even for weak SC coupling. In Fig.\ \ref{fig-dis} we show 
the amplitude and phase spectral functions at $\bq=0$ computed at RPA level\cite{suppl} for two indicative values of SC coupling and disorder. As one can see, the Higgs  mode acquires spectral weight below $2E_{gap}$ already for $V_0/t=0.5$ (upper panels), where $E_{gap}$ and $\D_0$ still coincide. 
For $U/t=5$ this sub-gap feature closely follows the one seen in the phase sector, where it can be interpreted as   a disorder-induced broadening of the sound mode, with a transfer of spectral weight from zero to finite frequency. Since the mixing between the amplitude and phase modes vanishes as $\o\ra 0$, see Eq.\ \pref{gaussian}, this feature appears as a finite-energy maximum in the amplitude sector.
At larger disorder $V_0/t=3$ (lower panels) the mixing of the amplitude with the phase is stronger at all energies, and also in this case the finite spectral weight of the Higgs mode below $2E_{gap}$  does not identify a well-defined resonance at the typical scale 2$\langle \D_0\rangle$. 

\begin{figure}[thb]
\includegraphics[width=7cm,clip=true]{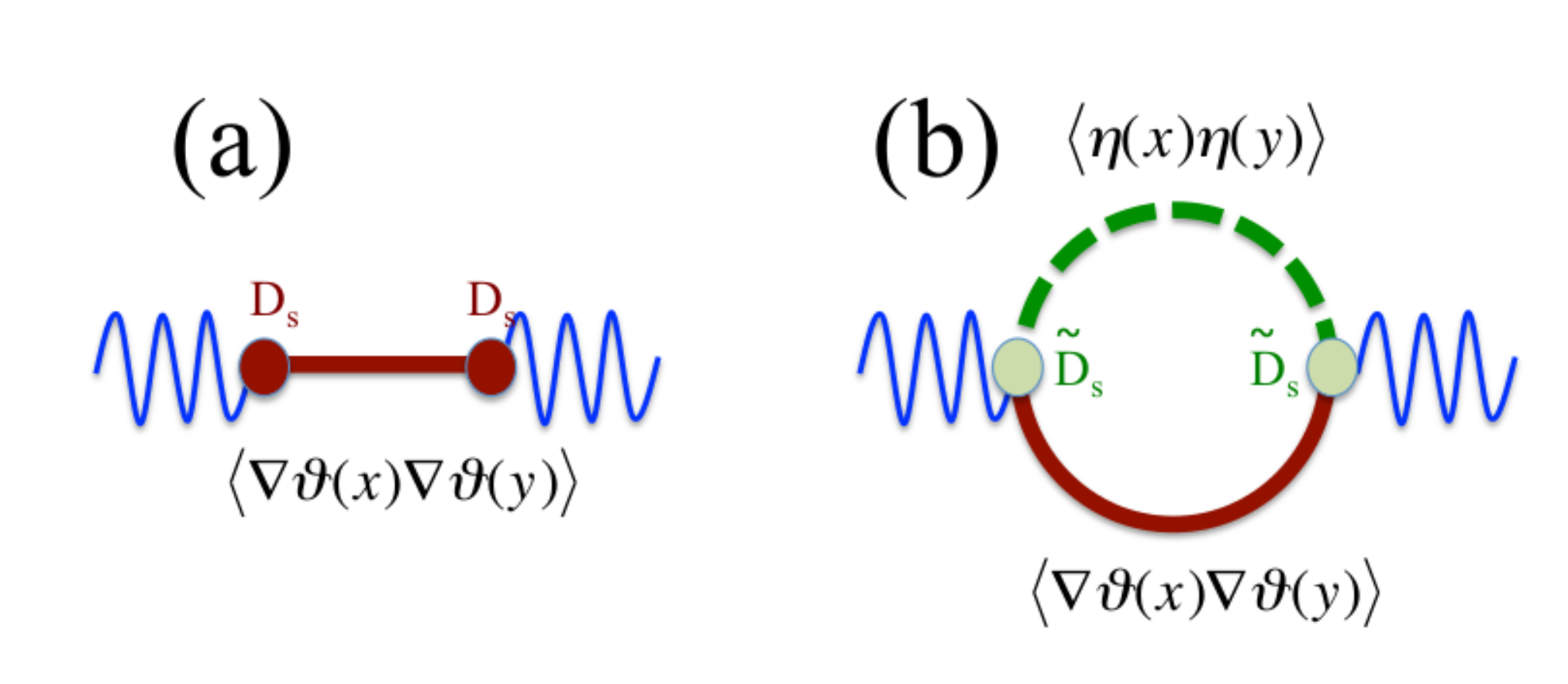}
\caption{Leading-order contributions of the SC collective modes to the conductivity, from the current given in Eq.\ \pref{curr}. Wavy, solid and dashed lines represent insertion of the electromagnetic field, the SC phase propagator and the SC amplitude propagators, respectively\cite{auerbach_prb10,sachdev_prb97}.}
\label{fig3}
\end{figure}

Let us finally comment on the relevance of these results for  experiments in disordered films of superconductors, where 
a dissipative absorption $\s(\o)$ at $\omega<2E_{gap}$ and $T\ll T_c$ has been recently  found\cite{armitage_prb07,frydman_natphys15,bachar_jltp14,practh_cm15}. Since in this regime quasiparticle excitations are suppressed, the extra-absorption should arise  from the SC collective modes, whose leading-order contribution to the current $\bJ$ can be written in general as\cite{suppl,nota_curr}
\be
\lb{curr}
{\bf J}=-2e \left(D_s \nb \theta+ 2\tilde D_s \eta\nb \theta \right), \quad \eta=\Delta/\D_0.
\ee
Since the optical conductivity is determined by the current-current correlation function, the leading-order optical process is proportional to  $~\langle \nb \theta (x) \nb \theta(y)\rangle$, i.e. to the excitation of a single phase mode (phason), see Fig.\ \ref{fig3}a, while at  higher order the convoluted process between one phason plus one Higgs mode is also possible, see Fig.\ \ref{fig3}b. In the clean case the one-phason process (a) contributes only to the delta-like superfluid response at $\omega=0$, while the process (b) contributes in general to $\sigma(\omega)$ at finite frequency, with an intensity and shape that depends on the coupling $\tilde D_s$ of Eq.\ \pref{curr} and on the spectral function of the Higgs mode. More specifically,  in the clean bosonic model \pref{so2}, where $\tilde D_s=D_s=c^2\D_0^2$ and the Higgs mode \pref{paradigm} has a sharp resonance at $\o=m$, the process (b) leads\cite{auerbach_prb10,podolsky_prb11,sachdev_prb97} to a finite-frequency optical absorption with an edge exactly at $m$.  By applying this result to superconducting films 
the authors of Ref.\ [\onlinecite{ frydman_natphys15}] interpreted the sub-gap optical  absorption found at strong disorder as an indirect evidence of the existence of a sharp  Higgs mode at an energy $m<2E_{gap}$. 
However, our calculations do not support this conclusion, since we explicitly demonstrated that for a realistic fermionic model with disorder  the spectral function of the Higgs mode does {\em not} identify any sub-gap LI resonance. In addition, we also checked\cite{suppl} that in the fermionic model \pref{H_MODEL} is  $\tilde D_s\simeq 0$ at weak coupling, strongly suppressing the 
optical process shown in Fig.\ \ref{fig3}b. 

 
  %
On the other hand, in the presence of disorder the one-phason process of Fig.\ \ref{fig3}a contributes also to the finite-frequency optical response\cite{stroud_prb00,randeria14,cea_prb14}, making it the most plausible candidate to explain the experimental findings\cite{armitage_prb07,frydman_natphys15,bachar_jltp14,practh_cm15}. In particular, for the fermionic model \pref{H_MODEL}  it has been proven\cite{cea_prb14} that the phase fluctuations give the largest contribution to sub-gap absorption, via a process equivalent to the one depicted in Fig.\ \ref{fig3}a. However, due to the non-trivial structure of the effective optical dipole of the phase mode, there is no simple correspondence between  its spectral function and  the optical conductivity, as one can see in Fig.\ \ref{fig-dis} where we report for comparison $\s(\o)$ from Ref.\ \cite{cea_prb14}. Note that
the optical conductivity is the response to the local, i.e. {\em screened} electric
field, and it is therefore determined by the current correlations functions irreducible with respect to the Coulomb interaction
\cite{cea_prb14}. In other words, what contributes to $\sigma(\o)$ are the SC collective mode of the neutral case, which are the ones shown in Fig.\ \ref{fig-dis}.
 
In summary, we studied the evolution of the amplitude (Higgs) mode in a lattice model for fermions. We showed that even when $2\D_0$ goes below the 
threshold $2E_{gap}$ for quasiparticle excitations the Higgs mode never identifies a sharp resonance at $2\D_0$. Indeed, the separation between $E_{gap}$ and $\D_0$, needed to remove the dynamical overscreening of quasiparticles to the Higgs mode, is only achieved by  an explicit breaking of the particle-hole symmetry, which implies a strong mixing of the amplitude and phase fluctuations. Thus, any sub-gap feature in the Higgs mode 
 is a signature of the underlying phase mode,  and it does not  resembles the sharp LI Higgs resonance found in the bosonic model \pref{so2}. These results establish that the outcomes of the relativistic $O(2)$ model \pref{so2} do not describe in general the physics of
 fermionic superconductors even in the "bosonic" limit, as it has been sometimes suggested\cite{auerbach_prb10,podolsky_prb11,frydman_natphys15}. On the other hand, the relativistic nature of the Higgs mode can be recovered in the BCS limit when a CDW gap contributes to make $E_{gap}$ larger than $\Delta_0$, as shown in Ref.\ \cite{cea_cdw_prb14}.  An interesting open question is the possibility that a similar mechanism can hold also in systems like cuprate superconductors, where increasing experimental evidence\cite{tranquada_science12,ghiringhelli_science12,hayden_natphys12,yazdani_science14} has been accumulating for the coexistence of SC and CDW order.
  
\vspace{0.5cm}  
  
We acknowledge useful discussions with J. Lorenzana.  This work has been supported  by Italian MIUR under projects FIRB-HybridNanoDev-RBFR1236VV, PRINRIDEIRON-2012X3YFZ2 and Premiali-2012 ABNANOTECH, and by the  Deutsche
Forschungsgemeinschaft under SE806/15-1.

%

\vspace{1cm}

\pagebreak
\clearpage

\onecolumngrid
\begin{center}
\textbf{\Large Supplemental Material}
\end{center}
\vspace{1cm}
\twocolumngrid

\setcounter{equation}{0}
\setcounter{figure}{0}
\setcounter{table}{0}
\setcounter{page}{1}
\makeatletter
\renewcommand{\theequation}{S\arabic{equation}}
\renewcommand{\thefigure}{S\arabic{figure}}
\renewcommand{\bibnumfmt}[1]{[S#1]}
\renewcommand{\citenumfont}[1]{S#1}

\section{Effective action for the collective modes: general formalism}
Let us start from Eq. (3) of the manuscript, that we report here:
\begin{equation}\lb{H_MODEL}
H=-t\sum_{\langle i,j\rangle\sigma}\left(c^\dagger_{i\sigma}c_{j\sigma}+h.c.\right)-U\sum_ic^\dagger_{i\uparrow}c^\dagger_{i\downarrow}c_{i\downarrow}c_{i\uparrow}.
\end{equation}
To investigate the physics of the collective modes around the mean field solution we follow the usual Hubbard-Stratonovich procedure\cite{nagaosa}, as implemented e.g. in Refs. \onlinecite{depalo_prb99,benfatto_prb04}. We then introduce in the action for the fermions a bosonic complex field $\psi_\D(\tau)$ which decouples the onsite interaction term of \eqref{H_MODEL} in the pairing channel. At $T<T_c$ one can choose to represent the superconducting (SC) fluctuations both in polar (amplitude and phase) or cartesian (real and imaginary parts) coordinates. In the former case the additional use of a Gauge transformation on the fermionic operators makes the dependence of the effective action on the time and spatial derivatives of the SC phase explicit, and it is then more convenient in the long-wavelength limit. The equivalence between the two approaches is guaranteed by the Ward identities, as discussed e.g. in Ref.\ [\onlinecite{marciani_prb13}]. We will then decompose $\psi_\D(\tau)=[\Delta_0 +\Delta_i(\tau)]e^{i\theta_i(\tau)}$, where $\Delta_i(\tau)$ represent the amplitude fluctuations of $\psi_\D$ around the mean-field value $\Delta_0$ and $\theta$ its phase fluctuations, which appears explicitly in the action after a Gauge transformation $c_i\ra c_ie^{i\theta_i/2}$. The interaction term of Eq.\ \pref{H_MODEL} can also be decoupled\cite{depalo_prb99,benfatto_prb04} in the particle-hole channel by means of a second (real) bosonic field $\psi_\rho=\rho_0+\rho$ which couples to the electronic density $\Phi_{\rho,i}=\sum_\sigma c^\dagger_{i\sigma}c_{i\sigma}$ and represents the density fluctuations $\rho$ of the system around the mean-field value $\rho_0$.

After the Hubbard-Stratonovich decoupling the action is quadratic in the fermionic fields that can then be integrated out leading to the effective action for the fields $\Delta$, $\theta$ and $\rho$:
\begin{equation}
\lb{seff}
	S_{eff}[\Delta,\theta,\rho]=S_{MF}+S_{FL}[\Delta,\theta,\rho]\quad,
\end{equation}
where $S_{MF}=\frac{N\Delta_0^2}{TU}+\frac{N\rho_0^2}{TU}-\text{Tr}\ln(-G_0^{-1})$ is the mean field action, 
$G_0^{-1}=i\o_n\hat\s_0-\xi_\bk\hat\s_3+\D_0\hat\s_1$ is the BCS Green's function and 
\be 
\lb{sfl}
S_{FL}=\sum_{n\ge1}\frac{\text{Tr}(G_0\Sigma)^n}{n}
\ee
is the fluctuating one, with the trace acting both in spin and momentum space, where $\Sigma_{kk'}$ denotes the self-energy for the fluctuating fields, which reads explicitly:
\begin{widetext}
\bea
\Sigma_{kk'}&=&
-\sqrt{\frac{T}{N}}\Delta(k-k')\sigma_1-\sqrt{\frac{T}{N}}\rho(k-k')\sigma_3-
\sqrt{\frac{T}{N}}\frac{i}{2}\theta(k-k')\left[
(k-k')_0\sigma_3-(\xi_\bk-\xi_{\bk'})\sigma_0
\right]-\nn\\
\lb{SELF_ENERGY}
&-&\frac{T}{2N}\sum_{q_i,\nu}\theta(q_1)\theta(q_2) \frac{\pd^2 \xi_\bk}{\pd k^2_\nu} \sin\frac{\bq_{1,\nu}}{2}\sin\frac{\bq_{2,\nu}}{2}
\sigma_3 \delta(q_1+q_2-k+k')+O\left(\theta^3\right)\text{ },
\eea
\end{widetext}
with $k=(i\Omega_n,\mathbf{k})$ and $\Omega_n=2\pi Tn$ bosonic Matsubara frequencies. Notice that the last line of Eq.\ \pref{SELF_ENERGY} represents the transcription on the lattice of the usual $(\nb \theta)^2$ term for a continuum model. In addition, in contrast to the continuum model, the lattice self-energy \pref{SELF_ENERGY} depends in principle\cite{depalo_prb99,benfatto_prb04} on all higher-order powers of the $\theta$ field, which are however irrelevant for the derivation of the Gaussian action. 

To derive the Gaussian action for SC fluctuations we should retain the terms up to $n=2$ in Eq.\ \pref{sfl}. The terms coming from an insertion 
of the $\sigma_0$ term of Eq.\ \pref{SELF_ENERGY} describe the effects of a paramagnetic current, so that they lead for example to the depletion of the superfluid stiffness at finite temperature due to quasiparticle excitations. On the other hand, mixed terms containing a $\hat \s_0$ times a $\hat \s_1,\hat\s_3$ matrix give higher-order contribution in $\bq$. Thus, since we are interested in the $T=0$ and long-wavelength limit, in the clean case we can safely neglect these terms. With lengthy but straightforward calculations one can then show that at gaussian level $S_{FL}$ can be written as:
 \begin{equation}
 \lb{GAUSS_ACTION_TOT}
	S_{FL}[\Delta,\theta,\rho]=\frac{1}{2}\sum_q
\Psi^\dagger(q)
	\hat{M}_{FL}(q)
	\Psi(q)\quad,
\end{equation}
with $\Psi^T(q)=\begin{pmatrix}\Delta(q)& \theta(q) & \rho(q)\end{pmatrix}$ and:
\begin{equation}
\lb{matrix}
\hat{M}_{FL}=\begin{pmatrix}
	2/U+\chi_{\D\D}(q)
	&
	\frac{i \o}{2}\chi_{\rho\D}(q)&
	\chi_{\rho\D}(q)\\
	-\frac{i\o}{2}\chi_{\rho\D}(-q)&
	\frac{\o^2}{4}\chi_{\rho\rho}(q)+\frac{D_s}{4}w(\mathbf{q})&
	-	\frac{i\o}{2}\chi_{\rho\rho}(q)\\
		\chi_{\rho\D}(-q)&
			\frac{i\o}{2}\chi_{\rho\rho}(q)&
				2/U+\chi_{\rho\rho}(q)
	\end{pmatrix}
\end{equation}
Here $\chi_{ij}(q)\equiv \frac{T}{N_s}\sum_k\text{Tr}\left[
	G_0(k+q)\sigma_iG_0(k)\sigma_j
	\right]$
are the response functions computed at BCS level and evaluated in the zero temperature limit. Since in Eq.\ \pref{SELF_ENERGY} the insertion of a $\hat\s_1$ or $\hat\s_3$ Pauli matrix corresponds to a term proportional to the amplitude or to the density/phase fluctuations, respectively, we  made the correspondence $1\ra \D, 3\ra \rho$ in the notation for the BCS susceptibilities that appear as coefficients of the action \pref{GAUSS_ACTION_TOT}-\pref{matrix}.  They are explicitly given at $T=0$ by:
\begin{subequations}\label{BUBBLES}
\begin{equation}
\lb{cdd0}
	\chi_{\D\D}(q)=\frac{1}{N_s}\sum_\mathbf{k}\frac{E^+_\mathbf{k}+E^-_\mathbf{k}}{E^+_\mathbf{k}E^-_\mathbf{k}}\cdot\frac{E^+_\mathbf{k}E^-_\mathbf{k}+\xi^+_\mathbf{k}\xi^-_\mathbf{k}-\Delta_0^2}{(\o+i\d)^2-\left(	E^+_\mathbf{k}+E^-_\mathbf{k}	\right)^2},
\end{equation}

\begin{equation}
\lb{chird0}
	\chi_{\rho\D}(q)=\frac{\Delta_0}{N_s}\sum_\mathbf{k}\frac{E^+_\mathbf{k}+E^-_\mathbf{k}}{E^+_\mathbf{k}E^-_\mathbf{k}}\cdot\frac{\xi^+_\mathbf{k}+\xi^-_\mathbf{k}}{(\o+i\d)^2-\left(	E^+_\mathbf{k}+E^-_\mathbf{k}	\right)^2},
\end{equation}

\begin{equation}
	\chi_{\rho\rho}(q)=\frac{1}{N_s}\sum_\mathbf{k}\frac{E^+_\mathbf{k}+E^-_\mathbf{k}}{E^+_\mathbf{k}E^-_\mathbf{k}}\cdot\frac{E^+_\mathbf{k}E^-_\mathbf{k}-\xi^+_\mathbf{k}\xi^-_\mathbf{k}+\Delta_0^2}{(\o+i\d)^2-\left(	E^+_\mathbf{k}+E^-_\mathbf{k}	\right)^2},
\end{equation}
\end{subequations}
with $E^\pm_\mathbf{k}\equiv E_{\mathbf{k\pm q/2}}$ and $\xi^\pm_\mathbf{k}\equiv \xi_{\mathbf{k\pm q/2}}$.

The quantity $D_s$ (see also Eq.\ \pref{ds} below) is the phase stiffness, that appears as coefficient of the term $w(\mathbf{q})\equiv 4\sum_\nu\sin^2(q_\nu/2)\simeq \mathbf{q}^2$, so that in the hydrodynamic limit the inverse of the bare phase-fluctuation propagator is: $4\hat{M}^{22}_\rho(q)\simeq -\kappa q_0^2+D_s\mathbf{q}^2$, which defines the BCS sound velocity: $v^0_s=\sqrt{D_s/\kappa}$, where  $\kappa\equiv -\chi_{\rho\rho}(0)=\frac{\Delta_0^2}{N_s}\sum_\mathbf{k}E_\mathbf{k}^{-3}$ is the bare compressibility.
By means of the self-consistent equation for $\Delta_0$, i.e. $2/U=\sum_\bk 1/E_\bk$ one can also rewrite the $\hat{M}^{11}_{\rho}$ term as:
\bea
2/U+\chi_{\D\D}(\o,0)&=&\frac{1}{N_s}\sum_\mathbf{k}\frac{\o^2-4\Delta_0^2}{E_\mathbf{k}\left[(\o+i\d)^2-4E_\mathbf{k}^2\right]}=\nn\\
\lb{deff}
&=&(4\Delta_0^2-\o^2)F(\o)
\eea
that corresponds to the expression used in Eq. (4) of the manuscript. To describe the fluctuations only in the SC sector we can integrate out the density fluctuations $\rho$ in Eq.\ \pref{matrix}, so that the Gaussian action for SC fluctuations reads:
\begin{equation}\label{GAUSS_ACTION_RPA}
S_{FL}[\Delta,\theta]=\frac{1}{2}\sum_q
\begin{pmatrix}\Delta(-q)& \theta(-q)\end{pmatrix}
	\hat{M}(q)
	\begin{pmatrix}\Delta(q)\\ \theta(q) \end{pmatrix}\quad,
\end{equation}
with $\hat M$ given by Eq. (4) of the manuscript, i.e.:
\begin{equation}\lb{RPA_MATRIX}
\hat{M}(q)=
	\begin{pmatrix}
	2/U+\tilde\chi_{\D\D}(q)&
	\frac{i\o}{2}\tilde\chi_{\rho\D}(q)\\
	-\frac{i\o}{2}\tilde\chi_{\rho\D}(-q)&
	\frac{\o^2}{4}\tilde\chi_{\rho\rho}(q)+\frac{D_s}{4}w(\mathbf{q})
	\end{pmatrix}\quad,
\end{equation}
The integration of the density field is equivalent as usual to the RPA dressing of the BCS susceptibilities. In particular, one has that 
$\tilde \chi_{ab}\equiv \chi_{ab}-\frac{\chi_{a\rho}\chi_{\rho b}}{2/U+\chi_{\rho\rho}}$. By using the equivalence  $1/(2/U+\chi_{\rho\rho})\equiv \tilde \chi_{\rho\rho}/\chi_{\rho\rho}$ one is then left with Eq. (4) of the manuscript.


\section{Expansion around $2E_{gap}$ at strong coupling}
Let us investigate in detail the amplitude fluctuations at $\mathbf{q}=0$. From Eq.\ \pref{RPA_MATRIX} we see that at $\bq=0$ the RPA resummation  of the bubbles in the density channel factorizes out, so that the spectral function of the Higgs 
$\rho_{\Delta}(\omega)\equiv\frac{1}{\pi}\mathrm {Im} \{1/{X}_{\Delta\Delta}(\omega+i\d,\mathbf{q}=0)\}$ 
is determined by the frequency behavior of the function:
\begin{equation}
\lb{xdd}
	X_{\Delta\Delta}\equiv \frac{2}{U}+\chi_{\D\D}-\frac{\chi_{\rho\D}^2}{\chi_{\rho\rho}}.
\end{equation}
Let us then investigate its behaviour from weak to strong coupling. As we discuss in the main text, at strong coupling one can reach the bosonic limit $E_{gap}\simeq U/2\gg \D_0$ only away from half filling, by moving the chemical potential 
 $\mu\simeq -U(1-n)/2$ below the band edge. In the following we will consider for instance the case where $\mu<0$ goes below the band edge $-4t$ at strong coupling. Thus, by introducing $E_{max}=\sqrt{\D_0^2+(4t+|\mu|)^2}$ and $E_{min}=\sqrt{\D_0^2+(4t-|\mu|)^2}$ one has that at strong coupling $E_{gap}\equiv E_{min}$. If we also approximate the momentum integration in the Eqs.\ \pref{BUBBLES} with an energy integration over a constant density of stated $N_F$ we can put: 
\begin{subequations}
\begin{equation}
\lb{chidd}
\frac{2}{U}+ \chi_{\D\D}=\left(4\Delta_0^2-\omega^2\right)I_0(\omega),
\end{equation}
\begin{equation}
\lb{chirr}
\chi_{\rho\rho}(\omega)=4\Delta_0^2 I_0(\omega),
\end{equation}
\begin{equation}
\lb{chird}
\chi_{\rho\D}(\omega)=4\Delta_0 I_1(\omega),
\end{equation}
\end{subequations}
where  the real and imaginary parts of the $I_{0,1}$ functions are given explicitly by:
\begin{widetext}
\begin{subequations}
\bea
	\mathrm{Re}I_0(\omega)&=&-{\cal P}\int_{-\mu-4t}^{-\mu+4t}\frac{\,d\xi \, N_F}{\sqrt{\D_0^2+\xi^2}(\omega^2-4\Delta_0^2-4\xi^2)}=\\
	\lb{I0r}
	&=&\frac{N_F\Theta(2\D_0-\o)}{\omega\sqrt{4\Delta_0^2-\omega^2}} \left[ \arctan\left( \frac {(4t+|\mu|)\o}{\sqrt{4\Delta_0^2-\omega^2}E_{max}} \right)+\arctan\left( \frac {(4t-|\mu|)\o}{\sqrt{4\Delta_0^2-\omega^2}E_{max}} \right)
	\right]\\
\lb{I0rg}
&+&\frac{N_F\Theta(\omega-2\Delta_0)}{2\omega\sqrt{\omega^2-4\Delta_0^2}}
	\ln \left| \frac{(4t+|\mu|)\o+E_{max}\sqrt{\omega^2-4\Delta_0^2}}{(4t+|\mu|)\o-E_{max}\sqrt{\omega^2-4\Delta_0^2}}\cdot
	\frac{(4t-|\mu|)\o+E_{min}\sqrt{\omega^2-4\Delta_0^2}}{(4t-|\mu|)\o-E_{min}\sqrt{\omega^2-4\Delta_0^2}}\right|	
	\\
\lb{I0im}
\mathrm{Im}I_0(\omega)&=& \left\{
\begin{matrix}
\Theta(\omega-2\Delta_0)\frac{\pi}{\o\sqrt{\omega^2-4\Delta_0^2} }, \quad |\mu|<4t,\\
\Theta(\omega-2E_{min})\frac{\pi}{2\o\sqrt{\omega^2-4\Delta_0^2} }, \quad |\mu|>4t
\end{matrix}
\right.
\eea
\end{subequations}

\begin{subequations}
\bea
\lb{I1r}
	\mathrm{Re}I_1(\omega)&=& {\cal P}\int_{-\mu-4t}^{-\mu+4t}\frac{\,d\xi \, N_F \xi}{\sqrt{\D_0^2+\xi^2}(\omega^2-4\Delta_0^2-4\xi^2)}=\frac{N_F}{4\omega}\ln \left| \frac{\omega+2 E_{max}}{\omega-2 E_{max}}\cdot \frac{\omega-2 E_{min}}{\omega+2 E_{min}}\right|\\
\mathrm{Im}I_1(\omega)&=& -
\Theta(\omega-2E_{min})\frac{\pi}{4\o}, \quad |\mu|>4t
\eea
\end{subequations}

\end{widetext}
so that e.g.  $I_0$ is an approximated expression for the function $F(\o)$ introduced in Eq.\ \pref{deff} above. 

\begin{figure}
\centering
\includegraphics[scale=0.4]{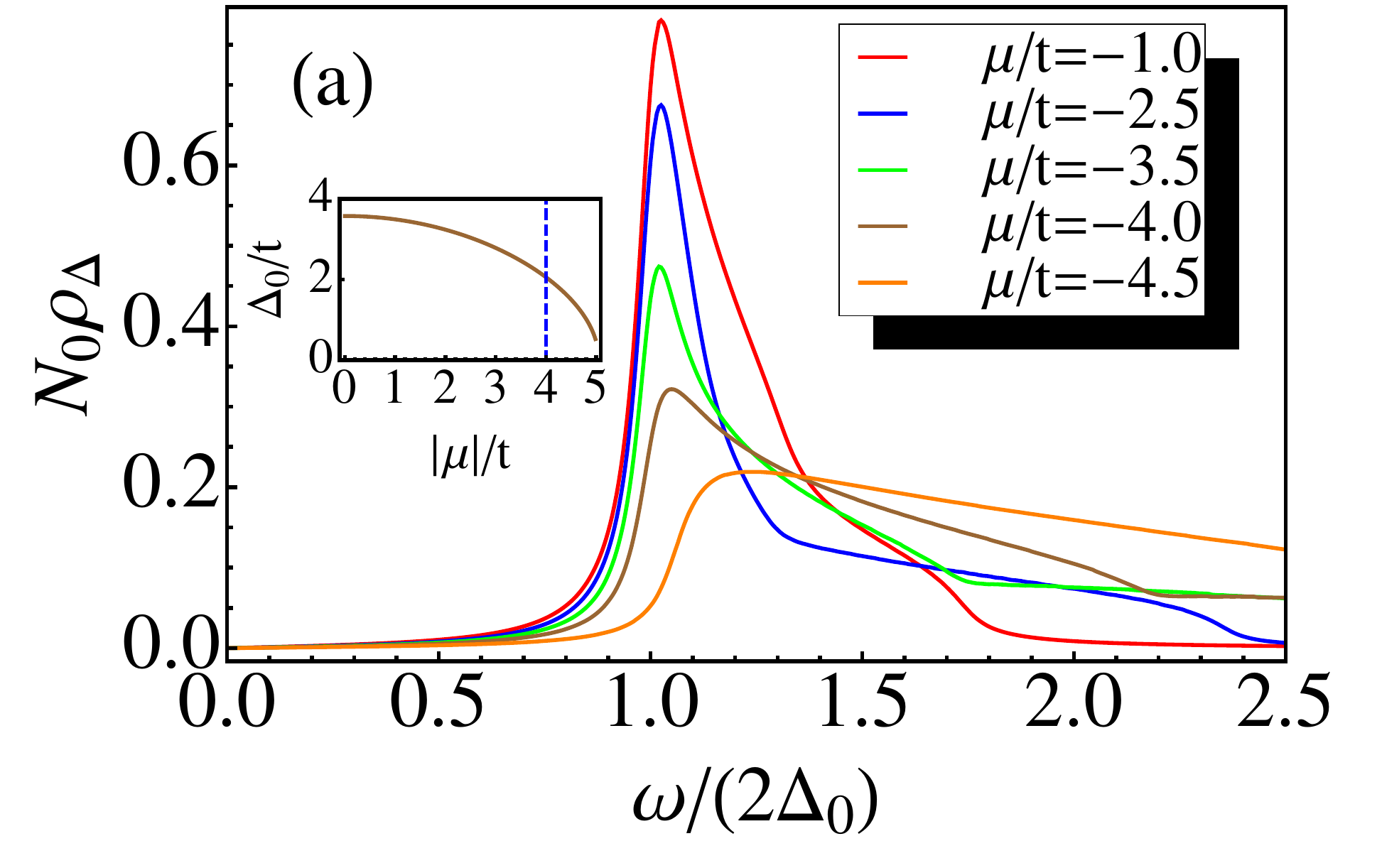}
\includegraphics[scale=0.38]{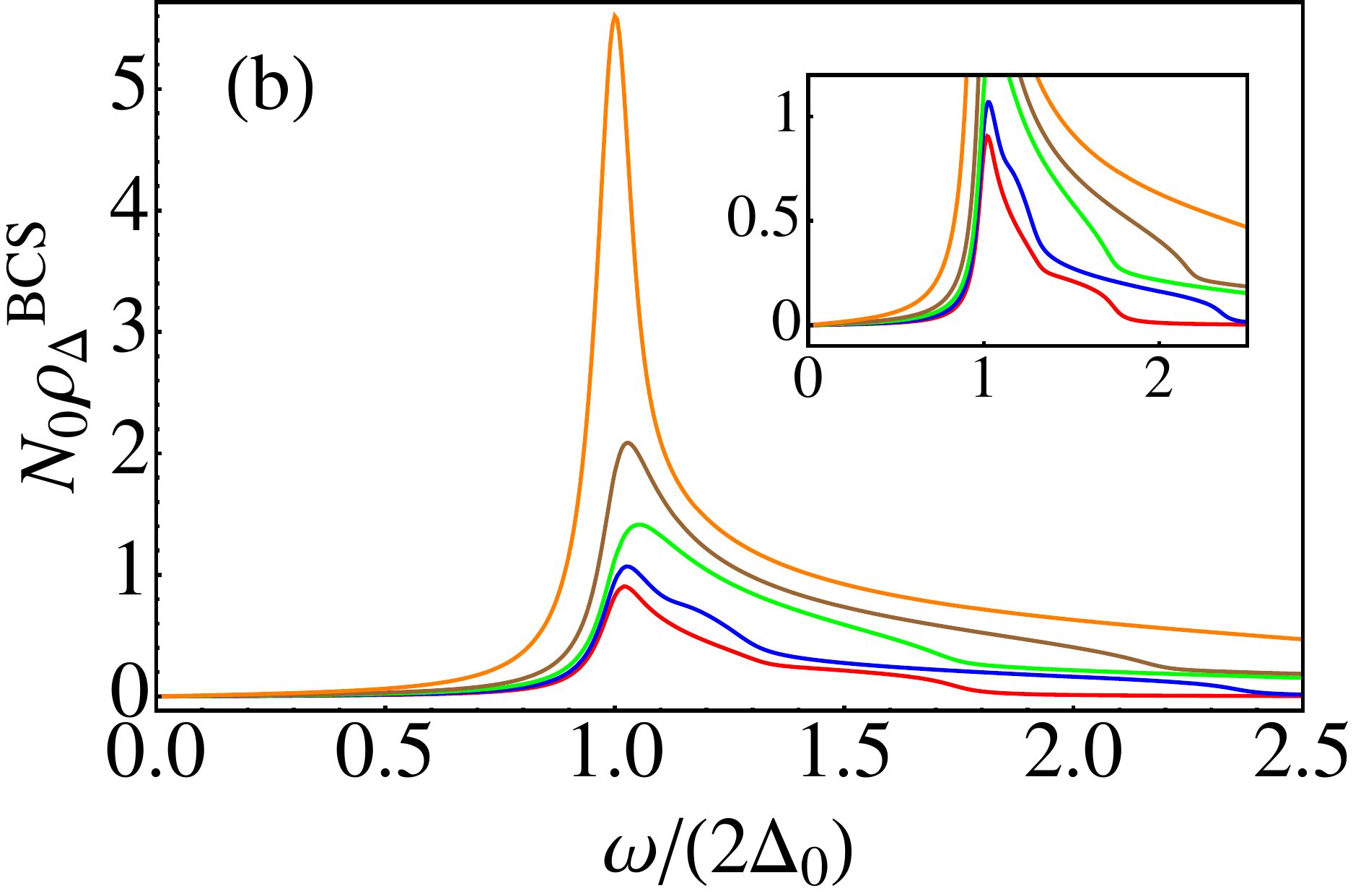}
\caption{(a) Frequency dependence of the spectral function for the amplitude fluctuations at $\bq=0$ at $U=8t$ for increasing value of the chemical potential. The numerical results are obtained with a broadening $\delta=0.1\D_0$. The scaling factor $N_0$ represents the equivalent density of states needed to match the numerical result with the analytical expressions discussed in the text, and it accounts for the differences due to the change of doping. The inset shows the evolution of the SC order parameter $\D_0$. For comparison we show in panel (2) the spectral function of the BCS term \pref{chidd} alone, that neglects the mixing to the phase sector. As one can see, as soon as the chemical potential moves outside the band edge ($|\mu|>4t$) the BCS spectral function in panel (b) displays a strong sharpening, since the quasiparticle contribution encoded in the function $I_0(\o)$ is no more diverging at $\o=2\D_0$, and the (bare) Higgs mode recovers a LI dynamics. However, as shown in panel (a), the coupling to the phase mode removes completely the signature at $2\D_0$ in the full Higgs spectral function and only a weak signature at $2E_{gap}>2\D_0$ is left. The inset of panel (b) is a zoom around $2\D_0$.}
\label{SPECTRAL_FUNCTION}
\end{figure}

In the weak-coupling regime the chemical potential lies in general well inside the band $|\mu|\ll 4t$, so that $E_{max}\simeq E_{min}$ and the 
integral \pref{I1r} is approximately zero.  This leads to the well-known decoupling of the amplitude and phase sectors due to the (approximate) particle-hole symmetry of the BCS solution, already mentioned in the main text. In addition, as far as $|\mu|<4t$ the two arctan in Eq.\ \pref{I0r} have the same sign as $\o\ra 2\D_0^-$, so that they give a constant contribution and $\mathrm{Re}I_0(\omega)$ diverges as $1/\sqrt{4\Delta_0^2-\omega^2}$, leading to $\mathrm{Re}X_{\Delta\Delta}(\omega)\simeq 
(\o^2-4\D_0^2)I_0(\o)\propto \sqrt{4\Delta_0^2-\omega^2}$ at the Higgs pole. At the same time $\mathrm{Re} I_0$ is finite as $\o\ra 2\D_0^+$ since the logarithmic term in Eq.\ \pref{I0rg} vanishes exactly as $\sqrt{\o^2-4\D_0^2}$, so that $\mathrm{Re}X_{\Delta\Delta}(\omega)\simeq 
(\o^2-4\D_0^2)$ as $\o>2\D_0$. On the other hand when $\o>2\D_0$ the imaginary part $\mathrm{Im}I_0(\o)$ diverges, see Eq.\ \pref{I0im}, so that $\mathrm{Im}X_{\D\D}$ grows as $\sqrt{\o^2-4\Delta_0^2}$ as $\o>2\D_0$, leading to the typical  over-damped resonance at $2\Delta_0$ in the spectral function $\rho_{\Delta}$, as shown in Fig.\ \ref{SPECTRAL_FUNCTION}a. 
It is worth noting that even retaining the small but finite value of $\mathrm{Re}I_1(\o)$ does not change these results. Indeed, as far as $\mu$ lies inside the band the function $I_1(\omega)$ (and then $\chi_{\rho\D}$) remains finite around $\omega=2\Delta_0$, see Eq.\ \pref{I1r}, while $\chi_{\rho\rho}$ is proportional to $I_0$, leading to the same results discussed so far in the case of perfect particle-hole symmetry. 

At strong coupling (and away from half filling) as soon as $|\mu|>4t$ the 
optical gap moves from $2\Delta_0$ to the higher value $2E_{gap}\equiv2 E_{min}$. In this case the two arctan in Eq.\ \pref{I0r} have {\em opposite} sign as $\o\ra 2\D_0^-$, so that they give overall a contribution proportional to $\sqrt{4\Delta_0^2-\omega^2}$ that removes the divergence of $\mathrm{Re}I_0(\omega=2\D_0)$. At the same time $\mathrm{Im}I_0$ is only finite above $2E_{gap}$, see Eq.\ \pref{I0im}, so that the amplitude fluctuations described at BCS level by the single \pref{chidd} term would recover a perfect Lorentz-invariant (LI) dynamics, with a sharp spectral function at $2\D_0$. However, as soon as $\mu$ moves outside the band edge the particle-hole symmetry is strongly violated and the coupling to the phase, dictated by the $\chi_{\rho\D}\propto I_1$ function, becomes large, moving the spectral weight of the Higgs fluctuations away from $2\D_0$ towards the new optical gap $2E_{gap}$. In particular one can see from Eqs.\ \pref{I0rg} and \pref{I1r} that both $\mathrm{Re}I_0$ and $\mathrm{Re}I_1$ diverge logarithmically as $\o\ra 2E_{min}\equiv 2E_{gap}$, so that:
\begin{equation}
\lb{I0as}
\mathrm{Re}	I_0(\omega)\simeq-\frac{N_F}{8E_{gap}(|\mu|-4t)}\left|\ln\left(1-\frac{\omega^2}{4E_{gap}^2}\right)\right|+K_0\quad;
\end{equation}
\begin{equation}
\lb{I1as}
\mathrm{Re}	I_1(\omega)\simeq -\frac{N_F}{8E_{gap}}\left|\ln\left(1-\frac{\omega^2}{4E_{gap}^2}\right)\right|+K_1\quad.
\end{equation}
where $K_0=\frac{1}{8E_{gap}(\mu-4t)}\ln\left[	\frac{\Delta_0^2}{4(\mu-4t)^2}\frac{(\mu+4t)E_{gap}+(\mu-4t)E_{max}}{(\mu+4t)E_{gap}-(\mu-4t)E_{max}}	\right]$ and $K_1=\frac{1}{8E_{gap}}\ln\left(4\frac{E_{max}-E_{gap}}{E_{max}+E_{gap}}\right)$.  According to Eqs.\ \pref{I0as}-\pref{I1as} now all the bubbles entering the definition of the inverse Higgs propagator \pref{xdd} are singular at $2E_{gap}$, so that at $\o\lesssim 2E_{gap}$:
\bea
\lb{asymp}
X_{\D\D}&=&  I_0(\o)\left [\o^2-4\D_0^2-4\left(\frac{I_1}{I_0}\right)^2\right]=\nn\\
&\simeq& 8(\mu-4t)^2K_0+8(\mu-4t)K_1
\eea
where we used the fact that $\o^2-4\D_0^2-4(I_1/I_0)^2\simeq \o^2-4E_{gap}^2$ (plus $1/|\ln(\dots)|$ terms) so that the divergence of $I_0(\o)$ as $\o\to 2E_{gap}$ is compensated by the quantity in square brackets in Eq.\ \pref{asymp}, and only a finite value remains. This result coincides with the simplified expression computed at $t=0$ and quoted in Eq. (10) of the main manuscript. The finite value of $\mathrm{Re} X_{\D\D}$ as $\o\to 2E_{gap}$, and the finite values of the imaginary parts of the $I_{0},I_{1}$ functions at $2E_{gap}$, imply  an even weaker resonance of the Higgs spectral function at the quasiparticle threshold $2E_{gap}$ when compared to the BCS case. This is clearly shown in Fig.\ \ref{SPECTRAL_FUNCTION}, where we report  the spectral function of the Higgs $\rho_{\Delta}(\omega)$ computed at $\bq=0$ with a numerical computation of the full expression \pref{xdd} on the lattice model. As one can see, the Higgs spectral function is always overdamped and, as $|\mu|$ exceeds the value $4t$, its maximum moves away from $2\Delta_0$ towards slightly higher frequencies, with a further weakening and broadening of the optical-gap signature. For comparison we also show in Fig. \ref{SPECTRAL_FUNCTION}(b) the BCS spectral function, i.e. the bare term \pref{chidd} in the inverse Higgs propagator, $\rho^{BCS}_{\Delta}(\omega)\equiv\frac{1}{\pi}{\mathrm{Im}}\{1/(2/U+\chi_{\D\D}(\o+i\d,\bq=0))\}$.
Here as soon as $\mu$ goes below the band edge the quasiparticle continuum moves away from the Higgs pole leading to a sharp LI resonance. However, the unavoidable mixing to the phase removes completely this signature, leading back to the broad spectral function shown in panel (a). 

\section{Role of long-range interactions}
The basic mechanism discussed in the manuscript (see also previous Section) leading to to the disappearance of the Higgs pole at $2\Delta_0$ in the strong-coupling case holds also in the presence of long-range interactions, which modify however the nature of the phase mode. The effect of Coulomb interactions can be taken into account by adding to the Hamiltonian \pref{H_MODEL} an interacting term
\be
\lb{hc}
H_c=\frac{1}{2}\sum_{\bk,\bk',\bq\atop{\sigma\sigma'}} V(\bq)c^\dagger_{\bk+\bq,\s}c^\dagger_{\bk'-\bq,\s'}c_{\bk',\s'}c_{\bk,\s}
\ee
where $V(\bq)$ is the Fourier transform of the Coulomb potential in the $D$ dimensional lattice. At small $\bq$ it reduces to the expression in the continuum limit, so that $V(\bq)\ra\lambda e^2/|\bq|^{D-1}$ where $\lambda=4\pi/\epsilon_B$ for $D=3$ while $\l=2\pi/\epsilon_B$ for $D=2$, $\e_B$ being the background dielectric constant. This term can be decoupled by means of the same HS field $\psi_\rho=\rho_o+\rho$ introduced in Eq.\ \pref{seff} to decouple the Hubbard term of Eq.\ \pref{H_MODEL} in the particle-hole channel. As a consequence, one can follow the same steps which lead to Eq.\ \pref{RPA_MATRIX} with the only difference that after integration of the density field the BCS bubbles are RPA dressed with the overall potential $U-2V(\bq)$. The matrix $\hat M$ matrix describing the coupled amplitude and phase fluctuations is then written as Eq.\ \pref{RPA_MATRIX}, corresponding to Eq.\ (4) of the main text:
\begin{equation}
\lb{gaussian}
\hat M=		\begin{pmatrix}
	(4\D_0^2-\o^2) F(\o) -\frac{(U-2V(\bq))\chi^2_{\rho\D}\tilde\chi^{LR}_{\rho\rho}}{2\chi_{\rho\rho}}&&
	\frac{i\o}{2} \frac{\chi_{\rho\D}\tilde\chi^{LR}_{\rho\rho}}{\chi_{\rho\rho}}\\
	-\frac{i\o}{2} \frac{\chi_{\rho\D}\tilde\chi^{LR}_{\rho\rho}}{\chi_{\rho\rho}} &&
	\frac{\o^2}{4}\tilde\chi^{LR}_{\rho\rho}+\frac{D_s}{4}\mathbf{q}^2
	\end{pmatrix}
	\end{equation}
where the dressed charge susceptibility $\tilde\chi^{LR}_{\rho\rho}$ reads:
\be
\tilde\chi^{LR}_{\rho\rho}\equiv \frac{\chi_{\rho\rho}}{1+\left(\frac{U}{2}-V(\bq)\right)\chi_{\rho\rho}}
\ee
If we now consider the long-wavelength limit we see that $\tilde\chi^{LR}_{\rho\rho}\ra -1/V(\bq)\ra 0$, so that the matrix \pref{gaussian} reduces to:
\begin{equation}
\lb{decoupled}
\hat M=		\begin{pmatrix}
	(4\D_0^2-\o^2) F(\o) -\frac{\chi^2_{\rho\D}}{\chi_{\rho\rho}}&&
	0 \\
	0 &&
	-\frac{\o^2}{4V(\bq)}+\frac{D_s}{4}\mathbf{q}^2
	\end{pmatrix}
	\end{equation}
In the weak-coupling BCS limit where $\chi_{\rho\D}\simeq 0$ one then recovers the equivalent of Eq. (7) of the main text, with the difference that now the phase mode is lifted to the plasmon. In particular, by considering the $\bq\ra 0$ limit of the Coulomb potential one then finds that the plasmon dispersion is $\omega_P(q) \simeq \sqrt{(2\pi e^2D_s / \e_B )q}$ in $D=2$ and $\omega_P(q)\simeq \sqrt{4\pi e^2/\e_B}$ in $D=3$. 
On the other hand in the strong-coupling regime, where also $\chi_{\rho\D}$ becomes singular at $2E_{gap}$, the pole of the amplitude mode is shifted away from $2\Delta_0$ by the charge-amplitude coupling, and one recovers for $|\hat M|$ the equivalent of Eq.\ (10) of the manuscript:
\bea
|\hat M|&=&\left[\frac{\o^2}{4V(\bq)}-\frac{D_s}{4}\mathbf{q}^2\right]
\left[ (4\D_0^2-\o^2)F(\o)-\frac{\chi_{\rho\D}^2}{\chi_{\rho\rho}}\right]\nn\simeq \\
&=&
\frac{\bq^2}{\l e^2/\e_B}\left[{\o^2}-\o_P^2(q)\right]\
F(\o)[4(\D_0^2+\mu^2)-\o^2]\simeq\nn\\
\lb{strong}
&\simeq&\frac{2\bq^2}{U\l e^2/\e_B}\left[{\o^2}-\o_P^2(q)\right].
\eea
As a consequence, the Coulomb interaction only modifies the nature of the phase mode, but it does not change the mechanism responsible for the lack of a Higgs signature at $2\Delta_0$ in the case where $E_{gap}>\Delta_0$. 

\begin{figure}
\centering
\includegraphics[scale=0.6]{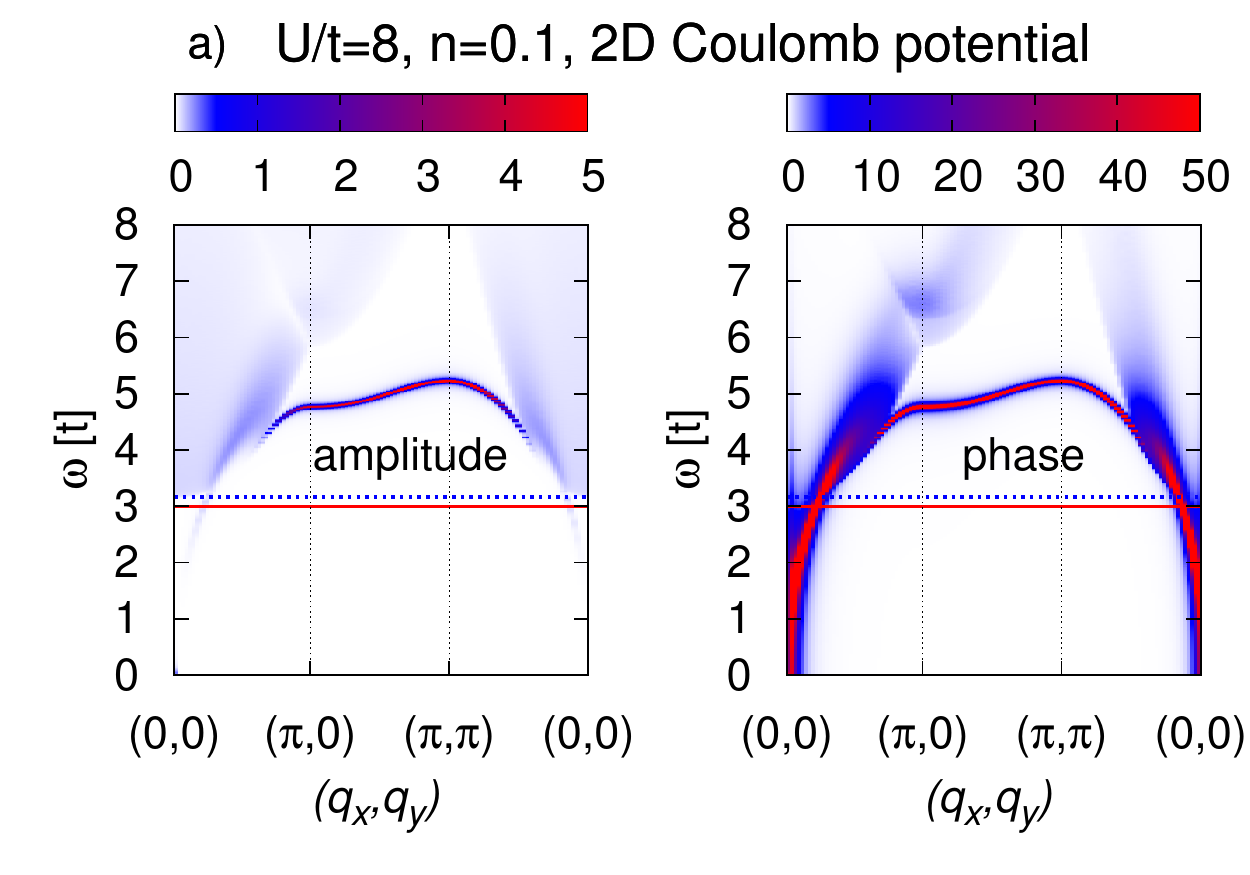}
\includegraphics[scale=0.6]{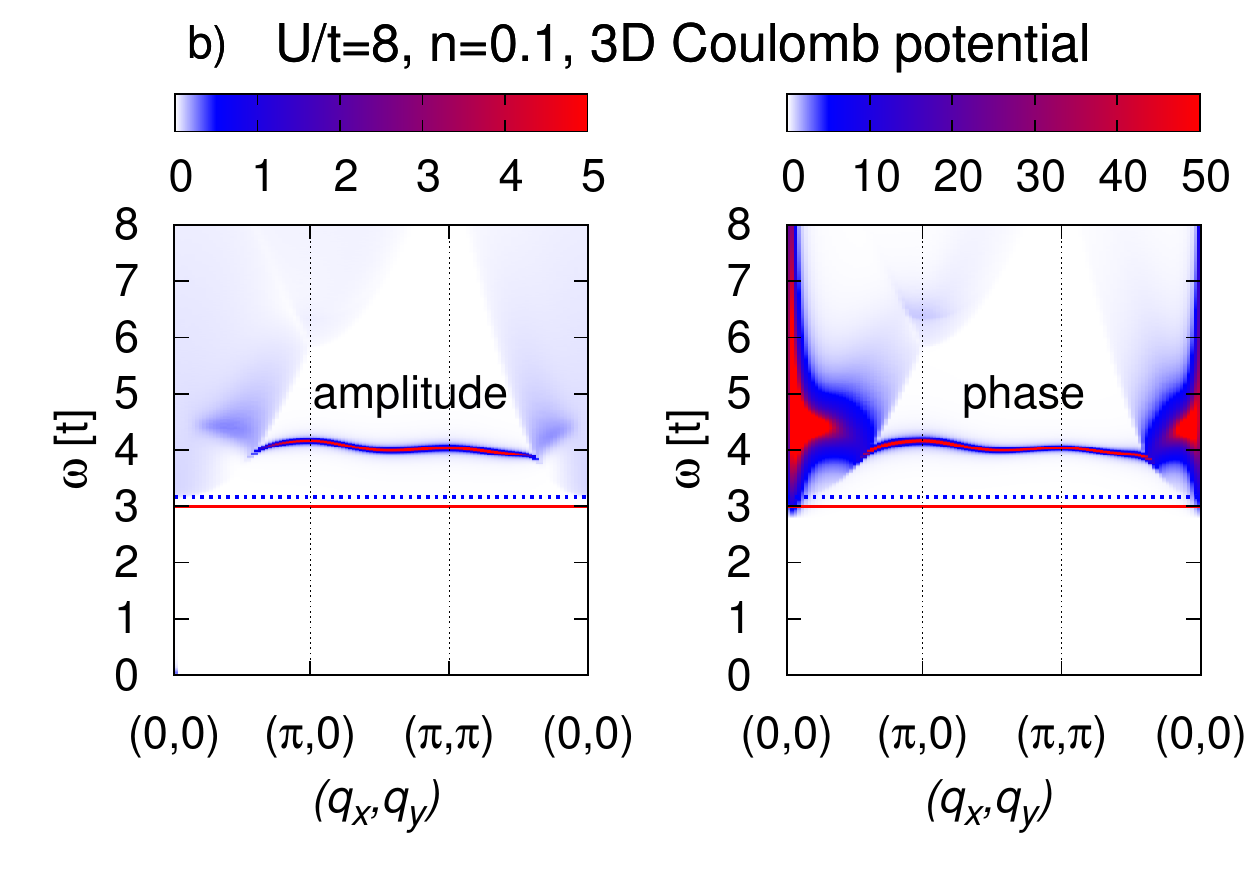}
\caption{
Intensity map of the spectral function of the amplitude and phase modes in the homogeneous case at  $U=8t$ for $n=0.1$ in the presence of a $2D$ (a) or $3D$ (b) Coulomb potential ($V_0/t=100$). The solid red lines mark the value of $2\Delta_0$ and the dashed blue lines the value of $2E_{gap}$.} 
\label{figcoul}
\end{figure}

These analytical estimates are confirmed by numerical computation of the amplitude and phase spectral functions shown in Fig.s \ref{figcoul}(a-b). Even though our model is in $D=2$ we nonetheless simulate also the case of a $3D$ plasmon for the sake of completeness. We adopt the following form
of the long-range Coulomb potential on a D-dimensional lattice
\begin{displaymath}
V^D(\bq)=\frac{V_0}{\left\lbrack2-\cos(q_x)-\cos(q_y)\right\rbrack^{(D-1)/2}}
\end{displaymath}
which at small momenta reproduces the $1/q^2$ ($1/q$) behavior in
$D=3$ ($D=2$) dimensions.
In the $2D$ case, see Fig.\ \ref{figcoul}a, the $\omega_P\sim \sqrt{q}$ dispersion of the plasmon found in the phase sector reflects also in the amplitude one at finite momentum. In the $3D$ case, see Fig.\ \ref{figcoul}b,  the plasma mode can be pushed even above $E_{gap}$ for sufficiently strong Coulomb potential, so that no spectral weight appears  below $2E_{gap}$. However, in both cases one clearly see that no signature is left at $2\Delta_0$, in full agreement with the conclusions of the manuscript, that holds irrespectively on the character (short or long range) of the density interactions. 

For what concerns the inhomogeneous case one  expects similar results: the presence of the Coulomb interactions does not eliminate the mixing between the amplitude and phase sector, but it affects in general the nature of the phase mode. On the other hand, since we are interested in making a comparison between the Higgs spectral function and the optical response, see Fig.\ 2 of the manuscript, we can in first approximation omit the RPA resummation of the Coulomb potential. Indeed, as we already discussed in Ref.\ [\onlinecite{cea_prb14}], the optical conductivity is the response
to the local, i.e. {\em screened}, electric field and therefore it is
determined by the irreducible current correlations with respect to the Coulomb potential. In other words, 
in disordered
systems the Coulomb interactions affect the phase mode, but
they do not affect its projection on the sub-gap optical absorption, whose understanding  is one of the main experimental motivations of our work.

\section{Spectral function in the disordered case}
In the Sec. I above we outlined the derivation of the effective action for the SC degrees of freedom by using a polar-coordinates representation for the
HS field $\psi_\D$ further supplemented by a Gauge transformation. Indeed, this approach makes more transparent the structure of the collective modes in the long-wavelength limit, discussed in details in the Sec. II. On the other hand, for the computation of the finite-$\bq$ amplitude and phase spectral function, and in particular for the evaluation of them in the presence of disorder, it is more convenient to introduce the pairing operators in real space:
\begin{equation}
\delta\eta_i \equiv c_{i\downarrow}c_{i\uparrow} - 
\langle c_{i\downarrow}c_{i\uparrow} \rangle \,,
\end{equation}
and to use a cartesian representation of  local amplitude ($A$) and phase 
($\Theta$) fluctuation operators
\begin{eqnarray*}
\delta A_i &\equiv & (\delta\eta_i+\delta\eta^\dagger_i)/\sqrt{2} \\
\delta \Theta_i &\equiv & i(\delta\eta_i-\delta\eta^\dagger_i)/\sqrt{2}\,.
\end{eqnarray*}
Together with the local charge fluctuation 
\begin{displaymath}
\delta \rho_i \equiv \sum_{\sigma}\left(c^\dagger_{i\sigma}c_{i\sigma} - \langle 
c^\dagger_{i\sigma}c_{i\sigma} \rangle\right)\,,
\end{displaymath}
one can set up a matrix of correlation functions 
\begin{equation}
\chi^{O,R}_{nm}(\omega)= i\int\!dt e^{i\omega t}
\langle {\cal T} \hat{O}_n(t) \hat{R}_m(0)\rangle
\end{equation}
which can be computed
on the RPA level as described in Appendix A of Ref. [\onlinecite{cea_prb14}].

For the homogeneous system one can compute explicitly the correlation functions in momentum space. With respect to the bare susceptibilities defined in Eqs. \pref{BUBBLES}  the only difference is in the one coupling the amplitude and phase sector, that reads:
\bea
& &\chi_0^{A\Theta}(q)=-\frac{1}{2N_s}\sum_\mathbf{k}
\left[ \frac{\xi_\bk^+}{E_\bk^+}+\frac{\xi_\bk^-}{E_\bk^-}\right]\times\nn\\
\lb{chiat}
& &\times \left[\frac{1}{\o+i\d-E_\bk^--E_\bk^+}+\frac{1}{\o+i\d+E_\bk^-+E_\bk^+}\right]
\eea
As one can see, as $\bq\ra $ one recovers $\chi_0^{A\Theta}=(\o/\D_0) \chi_{\rho\D}$, see Eq.\ \pref{chird0}. On the other hand at large $\bq$ to recover Eq.\ \pref{chiat} one should add in the expansion \pref{sfl} also the terms coming from the $\Sigma\sim \nb\theta \hat\s_0$ in Eq.\ \pref{SELF_ENERGY}, negligible in the long-wavelength limit. Apart from this, the RPA resummation (as discussed in Appendix A of Ref. \onlinecite{cea_prb14}) leads to the same scheme outlined in Sec. I. Indeed, at RPA level the present formulation in cartesian 
coordinates is completely equivalent to the polar representation of
SC fluctuations used in the previous sections, as one can show explicitly by means of the generalised Ward identities (cf. e.g. Ref.\ [\onlinecite{marciani_prb13}]). The imaginary parts of the full amplitude $\chi^{A,A}_{\bf q}(\omega)$ and phase
$\chi^{\Theta,\Theta}_{\bf q}(\omega)$ correlation functions are the quantities shown in Fig. 1 of the main paper.

To investigate the effect of disorder we add to Eq. \ref{H_MODEL} a random onsite potential,
\begin{equation}
H^{dis}=\sum_{i,\sigma} V_i c^\dagger_{i\sigma}c_{i\sigma} 
\end{equation}
with $-V_0 \le V_i\le V_0$ being a random variable which is taken 
from a flat and normalized distribution.
The correlation functions are then calculated on finite lattices
(typically $20\times 20$)  and averaged over disorder configurations. 
The same approach can be used to calculate the optical conductivity
in the presence of disorder (cf. Appendix A of Ref. [\onlinecite{cea_prb14}])
and the corresponding results are shown in Fig. 2 of the main manuscript.

\section{Superconducting collective-modes contribution to the current}

\begin{figure}[t]
\includegraphics[width=8cm,clip=true]{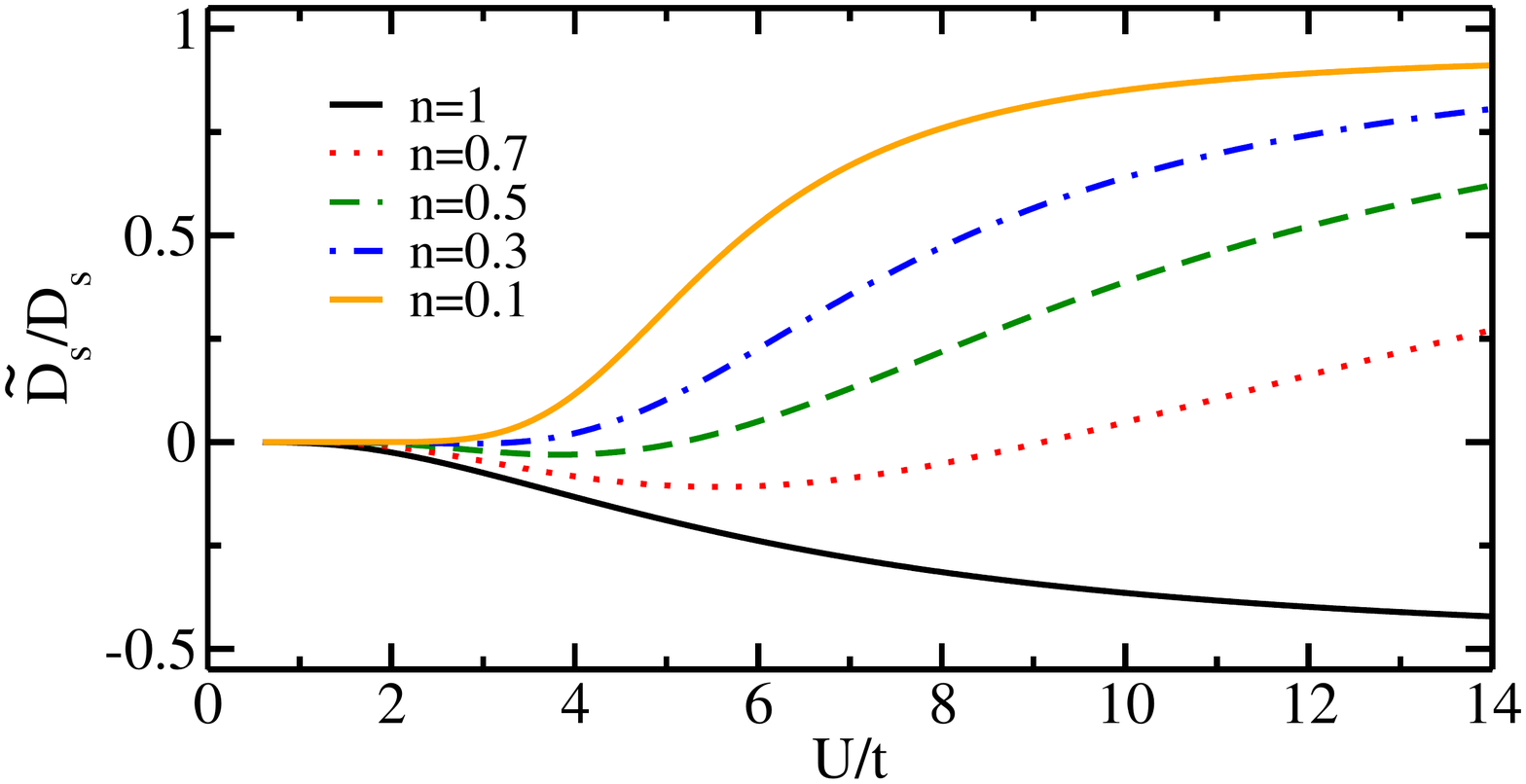}
\caption{Dependence on the SC coupling strength $U/t$ of the ratio $\tilde D_s/D_s$ appearing in the  definition \pref{curr} of the current for the homogeneous model \pref{H_MODEL} at various densities.}
\label{fig-ds}
\end{figure}

Let us outline briefly the derivation of Eq. (11) of the main manuscript. In the $O(2)$ model the current is easily obtained by means of the minimal-coupling substitution $-i\nb \ra -i\nb+2e\bA$ as
\bea
\bJ_s&=&-2e c^2\left(\psi\nb\psi^*-\psi^*\nb\psi\right)=\nn\\
&\simeq& -2e c^2 \D_0^2 (\nb\theta+2\eta\nb\theta)=\nn\\
\lb{curr}
&=&-2e( D_s\nb\theta+2\tilde D_s\eta\nb\theta)
\eea
where $D_s=\tilde D_s\equiv c^2\D_0^2$ and we put $\psi=\D_0(1+\eta)e^{i\theta}$. Here we retained only leading-order terms in the amplitude ($\eta$) and phase fluctuations, in analogy with the discussion of Refs. [\onlinecite{sachdev_prb97,auerbach_prb10,podolsky_prb11}]. The structure of the current in the lattice model \pref{H_MODEL} differs in general from the expression \pref{curr}. Indeed, apart from the presence of the quasiparticle contribution to the current, absent in the bosonic $O(2)$ model, the collective-modes contribution to the current does not display in general the local structure encoded in the form \pref{curr}. Nonetheless, since in general $\bJ_s=-\pd S_{FL}/\pd \bA$, a structure similar to Eq.\ \pref{curr} can be derived by considering in the 
action \pref{sfl} all the terms of the form:
\begin{equation}\label{O2_INV}
	S_{FL}\simeq \frac{1}{8}\int\,dx \left[D_s+2\tilde D_s\eta(x)\right]|\nabla\theta(x)|^2,
\end{equation}
where $D_s$ is the superfluid stiffness, defined as usual as:
\be
\lb{ds}
D_s=\frac{1}{2N_s}\sum_{\mathbf{k}\nu}\frac{\partial^2\xi_\mathbf{k}}{\partial k_\nu ^2}\left(1-\frac{\xi_\mathbf{k}}{E_\mathbf{k}}\right)
\ee

Let us then consider again the expansion \pref{sfl} and let us derive all the terms of third order in $S_{FL}^{(3)}=\text{Tr}\left[ G_0\Sigma^{(2)} G_0\Sigma^{(1)} \right]$ containing $(\nb \theta)^2$ times an other fluctuating field, that can be either the amplitude or the phase. By considering terms at second order $n=2$ in the action \pref{sfl} arising from the convolution between $\Sigma^{(2)}\simeq \theta^2 \hat \s_3$ and $\Sigma^{(1)} \simeq \Delta \hat \s_1+\rho \hat \s_3$ we have:
\bea
S_{FL}^{(3)}&=&\frac{1}{8}\int\,dx dy |\nabla\theta(x)|^2(\eta(y)+\rho(y)) R(x-y)=\nn\\
\lb{sfl3}
	&\simeq& \frac{1}{8}\int\,dx |\nabla\theta(x)|^2\left[ 2D_{\theta\Delta}\eta(x) +2D_{\theta\rho} \rho(x)\right]
\eea
where
\bea
\lb{dtd}
D_{\theta\Delta}&=&
		\frac{\Delta_0^2}{4N_s}\sum_{\mathbf{k}\nu}\frac{\partial^2\xi_\mathbf{k}}{\partial k_\nu^2}\frac{\xi_\mathbf{k}}{E_\mathbf{k}^3}=
		\frac{\Delta_0^2}{2N_s}\sum_{\mathbf{k}}\frac{\e_\bk \xi_\mathbf{k}}{E_\mathbf{k}^3}
		\\
\lb{dtr}
D_{\theta\rho}&=&\frac{\Delta_0^3}{4N_s}\sum_{\mathbf{k}\nu}\frac{\partial^2\xi_\mathbf{k}}{\partial k_\nu^2}\frac{1}{E_\mathbf{k}^3}=
\frac{\Delta_0^3}{2N_s}\sum_{\mathbf{k}}\frac{\e_\bk }{E_\mathbf{k}^3}
\eea
As one can see, in the continuum limit where $\xi_\bk \simeq \bk^2/2m-\mu$ one has that $D_s\simeq n/m$, $D_{\theta\D}\simeq  -(\D_0/2m)\chi_{\rho\D}(\o=0)$, and  $D_{\theta\rho}\simeq -(\D_0/2m)\chi_{\rho\rho}(\o=0)$, where $\chi_{\rho\D}$ and $\chi_{\rho\rho}$ are given by Eqs.\ \pref{chird} and \pref{chirr}, respectively. As a consequence, the approximate particle-hole symmetry of the BCS solution also guarantees that at weak coupling the phase gradient decouples from the amplitude fluctuations, while $D_{\theta\rho}$ is finite but usually smaller than $D_s$, since $D_{\theta\rho}\simeq (\D_0/m)N_F\simeq \D_0$ while $D_s$ is of order of the Fermi energy. In the opposite high-density limit, i.e. near half-filling where $\mu\simeq 0$, $D_{\theta\rho}\simeq 0$ while $D_{\theta\D}$ is finite, but always much smaller than $D_s$ since $D_{\theta\D}\sim \D_0 (\D_0/t) $ while $D_s\sim  t$, with $\D_0/t\ll 1$ at weak coupling. Thus, in the weak-coupling regime all the processes involving the excitation of one phason plus one Higgs or density mode are suppressed in the fermionic model \pref{H_MODEL}.

In the strong-coupling regime, where $\chi_{\rho\D}$ becomes as large as $\chi_{\rho\rho}$ away from half-filling, we expect that the coefficients \pref{dtd}-\pref{dtr} become as large as $D_s$, and a form similar to Eq.\ \pref{curr} can be recovered. To account also for the contribution of the density field, absent in the $O(2)$ bosonic model,  we will make the very rough approximation to consider its effect only in the renormalisation of the $D_{\theta\D} $ vertex of Eq.\ \pref{sfl3}, after integrating out the density field at RPA level. This procedure leads indeed to the form \pref{O2_INV} of the effective action, with the effective coefficient
\be
\lb{dts}
\tilde D_s=D_{\theta\Delta}-\frac{D_{\theta\rho}\chi_{\rho\D}(0)}{2/U+\chi_{\rho\rho}(0)}.
\ee
 In the strong-coupling regime the quantity \pref{dts} can be easily estimated at leading order in the small coupling $\a=(2t/U)$, see e.g. Ref.\ [\onlinecite{benfatto_prb02}]. In particular one has that:
\bea
\lb{crdas}
\chi_{\rho\D}&\simeq&-\frac{2}{U}\d\sqrt{1-\d^2}\\
\chi_{\rho\rho}&\simeq&-\frac{2}{U}(1-\d^2)\\
D_s&\simeq&U\a^2(1-\d^2)\\
D_{\theta\D}&\simeq&-\frac{U}{2}\a^2(1-\d^2)(1-3\delta^2)\\
D_{\theta\rho}&\simeq&\frac{3U}{2}\alpha^2\delta(1-\delta^2)^{3/2}
\eea
where $\d\equiv 1-n$ and Eq.\ \pref{crdas} only holds for $\delta\neq 0$.  As a consequence one immediately sees that  $\tilde D_s/D_s\ra 1$ for any finite doping away from half-filling, where the bosonic limit is recovered. On the other hand at half-filling $\chi_{\rho\D}=0$ at all values of $U$ so that $\tilde D_s=D_{\theta\D}$ always and $\tilde D_s/D\ra -1/2$ at strong coupling, as shown in Fig.\ \ref{fig-ds}.


\begin{thebibliography}{cc}
\bibitem{cg_note} 
In the phase-density sector one can also recover, at finite and large enough temperatures, a second mode called Carlson-Goldman\cite{cg_prl73} or Schmid-Sch\"on\cite{ss_prl75} mode,  as discussed e.g. in [\onlinecite{kulik81,
sharapov_cg_prb02}].
\end{thebibliography}
\end{document}